\documentclass[]{aastex631}

\usepackage{hyperref}
\shorttitle{CME Source Region Statistics}
\shortauthors{Majumdar et al.}
\graphicspath{{./}{figures/}}

\begin{document}

\title{A Coronal Mass Ejection Source Region Catalogue and Their Associated Properties}

\author[0000-0002-6553-3807]{Satabdwa Majumdar}
\affiliation{Aryabhatta Research Institute of Observational Sciences, Nainital, 263001, India}

\author[0000-0001-8504-2725]{Ritesh Patel}
\affiliation{Southwest Research Institute, 1050 Walnut Street, Suite 300, Boulder, CO 80302, USA}
\affiliation{Aryabhatta Research Institute of Observational Sciences, Nainital, 263001, India}

\author[0000-0002-6954-2276]{Vaibhav Pant}
\affiliation{Aryabhatta Research Institute of Observational Sciences, Nainital, 263001, India}

\author[0000-0003-4653-6823]{Dipankar Banerjee}
\affiliation{Aryabhatta Research Institute of Observational Sciences, Nainital, 263001, India}
\affiliation{Indian Institute of Astrophysics,2nd Block, Koramangala, Bangalore, 560034, India}
\affiliation{Center of Excellence in Space Science, IISER Kolkata, Kolkata 741246, India}

\author{Aarushi Rawat}
\affiliation{Department of Physics \& Astrophysics, University of Delhi, Delhi, 110007, India}

\author{Abhas Pradhan}
\affiliation{Department of Physics, Indian Institute of Technology Kharagpur, Kharagpur, 721302, India}


\author{Paritosh Singh}
\affiliation{UM-DAE Centre for Excellence in Basic Sciences, Mumbai, 400098, India.
}



\begin{abstract}

The primary objective of this study is to connect the coronal mass ejections (CMEs) to their source regions, primarily creating a CME source region (CSR) catalogue, and secondly probing into the influence the source regions have on different statistical properties of CMEs. We create a source region catalogue for 3327 CMEs from 1998 to 2017, thus capturing the different phases of cycle 23 and 24. The identified source regions are segregated into 3 classes, Active Regions (ARs), Prominence Eruptions (PEs) and Active Prominences (APs), while the CMEs are segregated into slow and fast based on their average projected speeds. We find the contribution of  these three source region types to the occurrences of slow and fast CMEs to be different in the above period. A study of the distribution of average speeds reveals different power-laws for CMEs originating from different sources, and the power-law is different during the different phases of cycles 23 and 24. A study of statistical latitudinal deflections showed equator-ward deflections, while the magnitude of deflections again bears an imprint of the source regions. An East-West asymmetry is also noted, particularly in the rising phase of cycle 23, with the presence of active longitudes for the CMEs, with a preference towards the Western part of the Sun. Our results show that different aspects of CME kinematics bear a strong imprint of the source regions they originate from, thus indicating the existence of different ejection and/or propagation mechanisms of these CMEs.
\end{abstract}


\section{Introduction} \label{intro}

With the brisk advancement in science and technology, especially with our dependencies on space satellites, radio communications, a good understanding of space weather has become more of a necessity than a luxury. In this context, coronal mass ejections (or CMEs) happen to be the major drivers of space weather and thus lie at the heart of our understanding of the same. These eruptions are capable of creating huge geomagnetic storms \citep{schwenn_2005,gosling_1993}. CMEs, in general are defined as discrete, bright, white light features that propagate outwards in the coronagraph field of view (FOV) from a timescale of few minutes to several hours \citep{hundhausen_1984}. This generic definition itself hints at the extent of contrast that is seen in the physical properties of CMEs, in terms of their masses, speeds, accelerations, widths and energies \citep{liv_rev_2012}. 

In order to study the behaviour of CMEs, they are usually tracked in the corona on images taken by a coronagraph, the obvious reason behind such practice rests on the ability of creating an artificial eclipse with the help of an occulting disk that blocks the disk of the Sun \citep{lyot_1930}. Unfortunately, this creation of an artificial eclipse, comes at the cost of obscuring the relevant disk features, and a reasonable part of the inner corona \citep[see Section 1 in ][]{majumdar_2020}. This would make the observer sometimes incapable of distinguishing a CME that originates around the disk center from a CME that originates at the limb \citep{majumdar_2021b}. As a consequence of this, primarily, a sense of the direction of propagation of the CME is lost, and secondarily, a rough estimate on the extent of projection effect that would certainly creep into the measurements, will also be lost \citep{balmaceda_2018}.  Thus it seems obvious that, white light coronagraphic images are not enough for a holistic understanding of CME properties and propagation, and that, it is equally important to locate and realise their source regions on the disk as well. This is also why extreme ultraviolet full disk imagers are extremely important in the study of CMEs. 

There have been several works in identifying and associating CMEs to their source regions. \cite{subramanian_2001} studied 32 CMEs from 1996 January to 1998 May, and they reported that $41\%$ of the CMEs were from Active Regions (ARs), $15\%$ were from quiescent prominences, while $44\%$ CMEs were connected to prominence eruptions that were connected to ARs. \cite{zhou_2003} studied 197 CMEs during the period from 1997 to 2001, and found that around $79\%$ of the CMEs originated from ARs, while the rest originated from outside ARs. There has also been several other efforts of studying the locations of CME sources on the solar disk \citep[][]{akiyama_2019,compagnino_2017,kim_2017,wang_2011,yashiro_2009,lara_2008,tripathi_2004}. \cite{gao_2014} reported that the association of solar surface activity with CMEs tend to vary during the rising and maximum phases of solar cycle 23 and 24. \cite{Moore_2007} showed that the width of the CME can be used to estimate the strength of the magnetic field of the source region. Recently, \cite{majumdar_2021} showed that the source regions of CMEs have a clear imprint on the coupling of CME kinematics happening at the inner and outer corona regions. It was also recently shown by \cite{pant_2021}, that the distribution of CME widths follow different power-laws for CMEs originating from different source regions, and a similar imprint is also seen in the evolution of the 3D volume of CMEs \citep{majumdar_2022}. Thus, it seems there is a close relationship between CME properties and the source region the CMEs originate from.

With regard to connecting the properties of CMEs to their source regions, a catalogue connecting the same have proven very useful over the past few years. One of the most extensively used CME catalogue is the Coordinated Data Analysis Workshop \citep[CDAW; ][]{cdaw} catalogue, which is a manual catalogue that detects and records different parameters of the CMEs, in the LASCO field of view (FOV). There has also been some automated or semi-automated catalogues like Solar Eruptive Event Detection System \citep[SEEDS; ][]{seeds}, Computer Aided CME Tracking \citep[CACTUS; ][]{pant_2016,cactus}, Automatic Recognition of Transient Events and Marseille Inventory from Synoptic maps \citep[ARTEMIS; ][]{artemis}, CORonal IMage Processing \citep[CORIMP; ][]{corimp_1,corimp_2}. A dual-viewpoint CME catalogue \citep[JHUAPL; ][]{jhuapl} on the other hand, relies on visual identification of CMEs, while the measurements are taken through a semi-automated algorithm. Although the above catalogues provide a comprehensive information of the CMEs that have been detected respectively, the one significant aspect that is missing in all of these catalogues, is the information of the source regions of all the CMEs that have been catalogued. 

In order to fill in this gap, we create a CME source region (CSR) \href{https://al1ssc.aries.res.in/catalogs}{catalogue} \footnote{\href {https://al1ssc.aries.res.in/catalogs}{https://al1ssc.aries.res.in/catalogs}},which is done by extracting CMEs (that occurred during the different phases of cycle 23 and 24) from the CDAW catalogue. We also study the impression of the identified source regions on different statistical properties of CMEs, which will further signify the importance of a source region catalogue for CMEs. In Section~\ref{data} we outline the different data sources that have been used in this work, followed by the working method that we have adopted. In Section~\ref{results}, we present the results our work and finally, in Section~\ref{conclusion}, we summarise the main conclusions of this work. 

\section{Data and method}

\subsection{Data Source} \label{data}

We use the CDAW catalogue to select and extract our events of interest. The source regions of the selected CMEs are identified by using the data from \textit{Extreme-Ultraviolet Imaging Telescope} \citep[EIT; ][]{delaboudini_1995} on-board \textit{Solar and Heliospheric Observatory}  \citep[SOHO; ][]{soho}, the {\textit{Atmospheric Imaging Assembly}} (AIA) \citep{aia} on-board {  \textit{Solar Dynamics Observatory}} (SDO), the two coronagraphs COR-1, COR-2  and the \textit{Extreme Ultraviolet Imager} \citep[EUVI; ][]{euvi} of the \textit{Sun Earth Connection Coro- nal and Heliospheric Investigation} \citep[SECCHI; ][]{howard_2008} on-board the \textit{Solar Terrestrial Relations Observatory} \citep[STEREO; ][]{stereo}, refer to Section for details on the source region identification.

\begin{figure*}[h]
\gridline{\fig{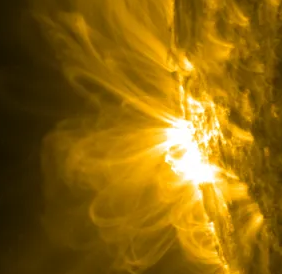}{0.3\textwidth}{(a) Active Region (AR)}
          \fig{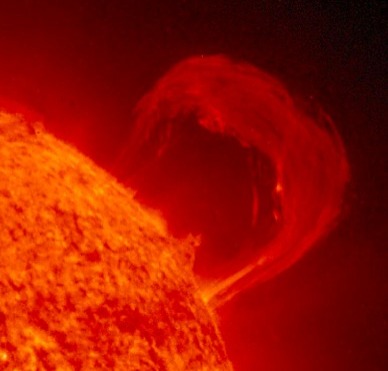}{0.3\textwidth}{(b) Prominence Eruption (PE)}
          \fig{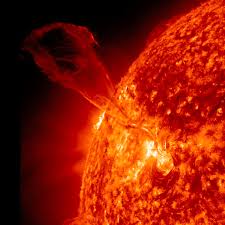}{0.3\textwidth}{(c) Active Prominence (AP)}
          }
\caption{This Figure shows examples of the three source region categories, that have been considered in this work. (a) An Active Region is shown, which is exhibiting bright emission in AIA 171 \AA, (b) shows a Prominence eruption (PE), which are the prominences located at the quiet Sun region is shown in AIA 304 \AA, and in (c) an Active Prominence in AIA 304 \AA, where we can see that the foot-point of the prominence is connected to an Active Region.
\label{cme_sr}}
\end{figure*}

\subsection{Event Selection} \label{event_select}

The CMEs for this work were selected from the CDAW catalogue in such a way that different phases of solar cycle 23 and 24 could be considered. Thus, we selected CMEs that occurred between January 1998 to December 2017. This allowed us to capture the rising phase, the maxima, and the declining phase of the cycles (Table~\ref{table1}). Firstly, the ``very poor" CMEs as quoted in the CDAW catalogue are removed from our sample, as \cite{wang_2014} had reported that the detection of these ``very poor" CMEs are based on the discretion of manual operators. Further, the source region identification and the derived kinematics of these ``very poor" CMEs is ambiguous. Thus in order to remove such untoward bias and any discrepancies, that might intrude in our analysis, we discard them.  Next, during the pre-STEREO era, on-disk observations of the backside of the Sun with respect to the Sun-Earth line were not available. This led to the removal of all those CMEs for which, no source region could be identified on the front-side of the Sun, as for such cases, the source regions were located behind the limb. Further, after the launch of STEREO, there were certain source regions which were located in the unobserved regions combining the STEREO and SOHO viewpoints, and thus it was not possible to identify their source regions either. During the period considered in this work, there were also certain cases of un-availability of data from the disk imaging instruments (mentioned in Section~\ref{data}), which further restricted the number of events.

\subsection{Classification of slow and fast CMEs}

The CMEs thus selected for the study are classified into slow and fast based on the average (linear) speed as provided in the CDAW catalogue. The CDAW catalogue in this regard, quotes the average speed at which the leading edge of the CME travels in the field of view of the coronagraphs. This classification is done based on the speed of the CME relative to the ambient solar wind speed. It is known that the slow solar wind usually travels with speeds less than around 400 km/s, and on the other hand, the fast solar wind shows evidence of speeds higher than 400 km km/s as reported by \cite{schwenn_2006}. Thus, for a statistical study, the average solar wind speed can be considered to be around 400 km/s. Based on this average solar wind speed, we tag CMEs with speeds higher than 500 km/s to be fast CMEs and the ones with speeds lesser than 300 km/s as slow CMEs \citep[see; ][]{pant_2021}. Further, since we are considering projected speeds for this classification, the CMEs with speeds ranging between 300 km/s to 500 km/s can neither be strictly considered fast, nor slow. This is due to the uncertainties in the projected speeds, and thus they are classified as intermediate CMEs. In this context, it should also be noted that since we are working with projected speeds, the results of this work will be suffering from projection effects, as the true speeds for the CMEs originating from the disk center will be different than their projected speeds.

\begin{figure*}[h]
\gridline{\fig{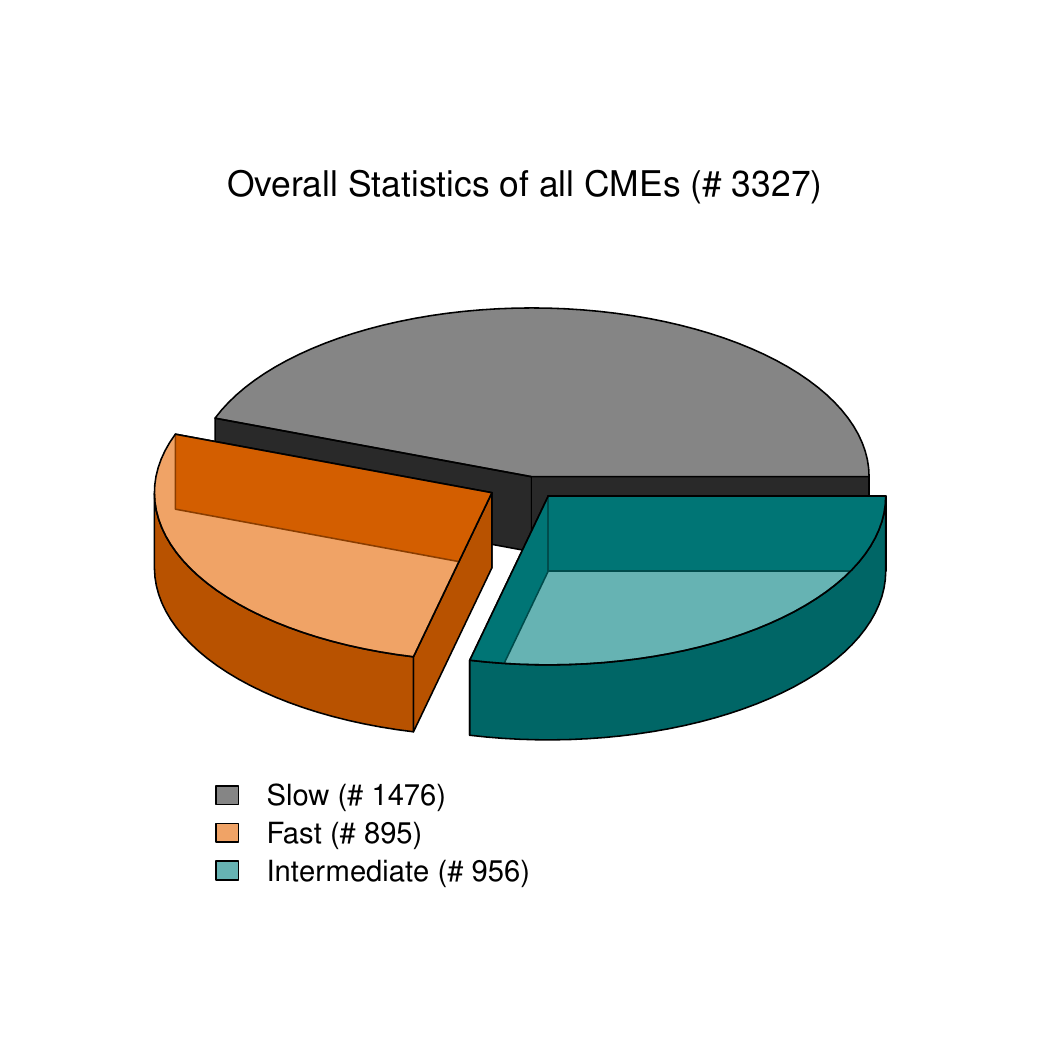}{0.45\textwidth}{(a)}
          \fig{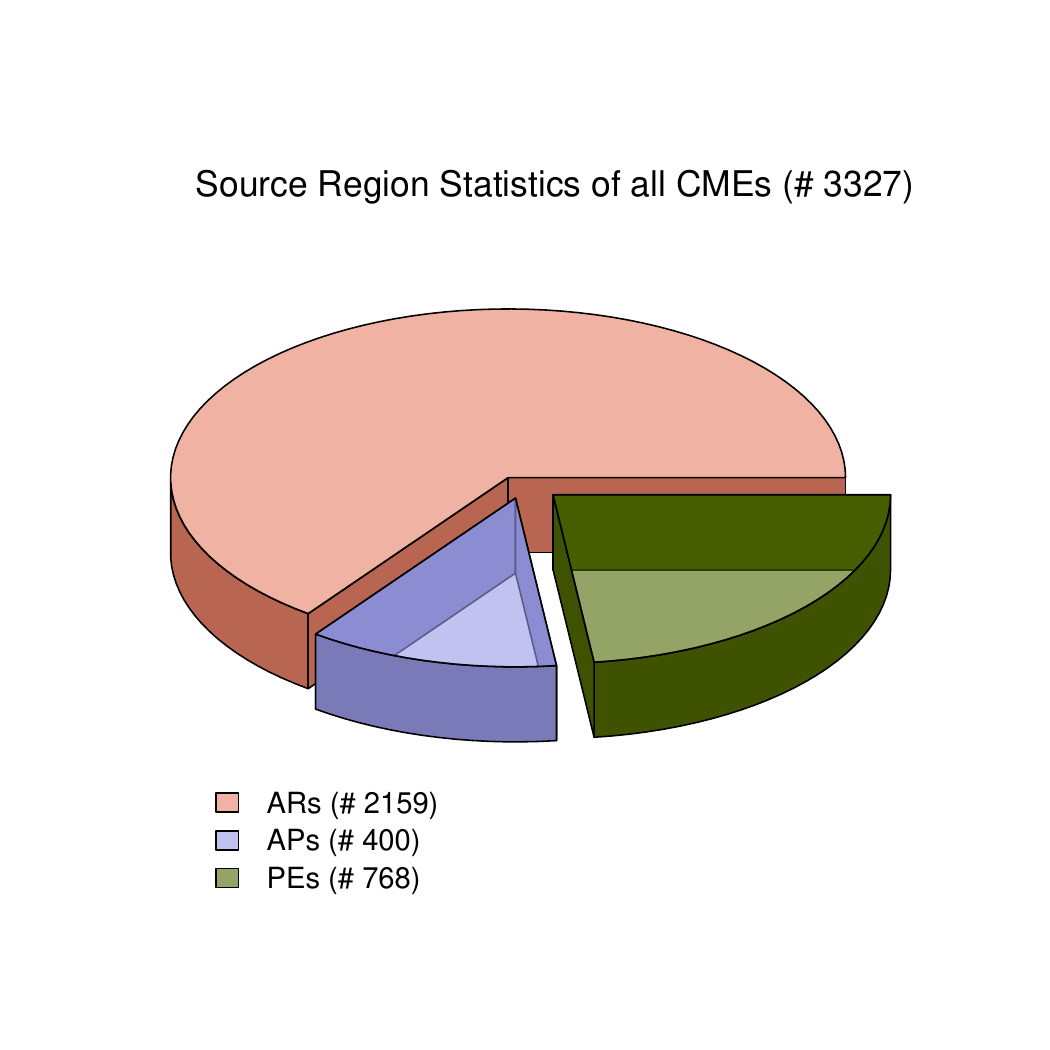}{0.45\textwidth}{(b)}
          }
\caption{Fractional occurrences of (a)Slow and fast CMEs and the fractional contribution of (b) different source regions to the CMEs during the studied period .
\label{all_cmes}}
\end{figure*}

\begin{center}
\begin{table}[]
    \centering
    \begin{tabular}{c|c|c|c}
    \hline \hline
    Solar Cycle & Cycle Phase & Year & No. Of Events  \\
    \hline \hline
     Cycle 23 & Rising & 1998 & 171\\
     Cycle 23 & Rising & 1999 & 174\\
        \hline
    Cycle 23 & Maxima & 2000 & 289\\
    Cycle 23 & Maxima & 2001 & 132\\
    Cycle 23 & Maxima & 2002 & 71\\
    \hline
    Cycle 23 & Declining & 2003 & 57\\
    Cycle 23 & Declining & 2004 & 218\\
    Cycle 23 & Declining & 2005 & 87\\
    \hline
    Cycle 23 & Minima & 2008 & 78\\
    Cycle 23 & Minima & 2009 & 113\\
    \hline \hline
    Cycle 24 & Rising & 2010 & 158\\
    Cycle 24 & Rising & 2011 & 269\\
    Cycle 24 & Rising & 2012 & 382\\
    \hline
    Cycle 24 & Maxima & 2013 & 860\\
    Cycle 24 & Maxima & 2014 & 23\\
    \hline
    Cycle 24 & Declining & 2017 & 245\\
           \hline \hline
    \end{tabular}
    \caption{The table shows the number CMEs studied in this work, during the different phases of cycle 23 and 24.} 
    \label{table1}
\end{table}
\end{center}

\subsection{Identification and Classification of Source Regions of CMEs}

For identifying and associating (spatially and temporally) the source regions of the CMEs, we follow a similar procedure as reported in \cite{pant_2021,majumdar_2020,subramanian_2001}, and we use the JHelioviewer software \citep{jhelio,jhelio2} to back-project the CMEs onto the solar disk. In order to do this, we consider a spatial and temporal window of criteria in order to associate a CME to its source region. For a spatial association between a source region and a CME originating from it, we require that the latitude of the source region to be around $\pm$ 30$^{\circ}$ to that of the PA of the center of the CME as reported in the CDAW catalogue \citep[see; ][]{pant_2021,majumdar_2020,Gilbert_2000}. When the spatial extent of the source region is large, the latitudes and longitudes of the boundaries are recorded and their mean is computed. For a temporal association, an eruption or radially outward motion is considered in the above mentioned window of latitude, in a minimum time of 30 minutes before the CME leading front is first observed in the LASCO C2 field of view. However, it must also be kept in mind that this temporal criteria was kept flexible and was decided based on the quoted speed of the CME in the CDAW catalogue. The different disk imaging and coronagraph data as mentioned in Section~\ref{data}, are loaded into the JHelioviewer software to create base difference movies that would help in connecting the outgoing CMEs to the source region on the disk. \\

After identifying the source regions by the above procedure, the identified source regions are then broadly classified into three categories: (a) Active Regions (ARs), (b) Prominence Eruptions (PEs) and (c) Active Prominences (APs) \citep[see; ][]{pant_2021,majumdar_2020}. The ARs are regions of predominantly strong magnetic field, that are much denser and hotter compared to the background plasma of the corona, producing bright emission in the extreme ultraviolet and soft X-ray wavelengths (Figure~\ref{cme_sr}(a)). Prominences are on the other hand, cool dense material (8000 K) which are embedded in the relatively hotter corona, which are identified as an emission feature when they are observed at the solar limb (Figure~\ref{cme_sr}(b)), and are identified as a dark absorption feature when they are seen on the disk, projected against the background hotter corona (called filament) \citep{Gilbert_2000}. A source region is considered  as a prominence eruption (PEs) if a strong radial motion is identified away from the surface of the Sun, where all or some part of the prominence material is observed to escape the Sun's gravitational field. For a filament eruption (which is included in the same category with PEs in this work), either a tangential motion across the Sun's surface was noted, followed by a subsequent eruption, or by simply noting any disappearance of a pre-existing filament, which is then followed by a transient coronal manifestation (also refer \citet{webb_1987}). Thus, it should be noted that PEs correspond to erupting prominences located in quiet Sun regions (outside ARs). The third category APs, considers PEs events where the prominence ( with their one or more foot-points) are embedded in ARs (Figure~\ref{cme_sr}(c)) \citep[see; ][]{subramanian_2001}. 

\section{Results}\label{results}

\subsection{Overall Statistics}

In Figure~\ref{all_cmes}(a), we show the overall fractional occurrences of slow, fast and intermediate CMEs during our studied period. We find that $\sim$43 $\%$ of the studied CMEs were slow CMEs, while $\sim$ 24$\%$ of the CMEs were fast CMEs and $\sim$ 33$\%$ were intermediate CMEs. Thus we see that a majority of our studied CMEs were slow CMEs. In Figure~\ref{all_cmes}(b), we show the overall contribution of the three different source regions that we have considered in this work. We find that ARs contribute to $\sim$65 $\%$ of the studied CMEs, the APs contribute to $\sim$12 $\%$ of the CMEs studied and $\sim$23 $\%$ of the total CMEs come from PEs. Thus, we find that although most of the CMEs are originating from ARs, there are also a considerable number of CMEs originating from PEs, which are located in the quiet Sun regions. 

\begin{figure*}[h]
\gridline{\fig{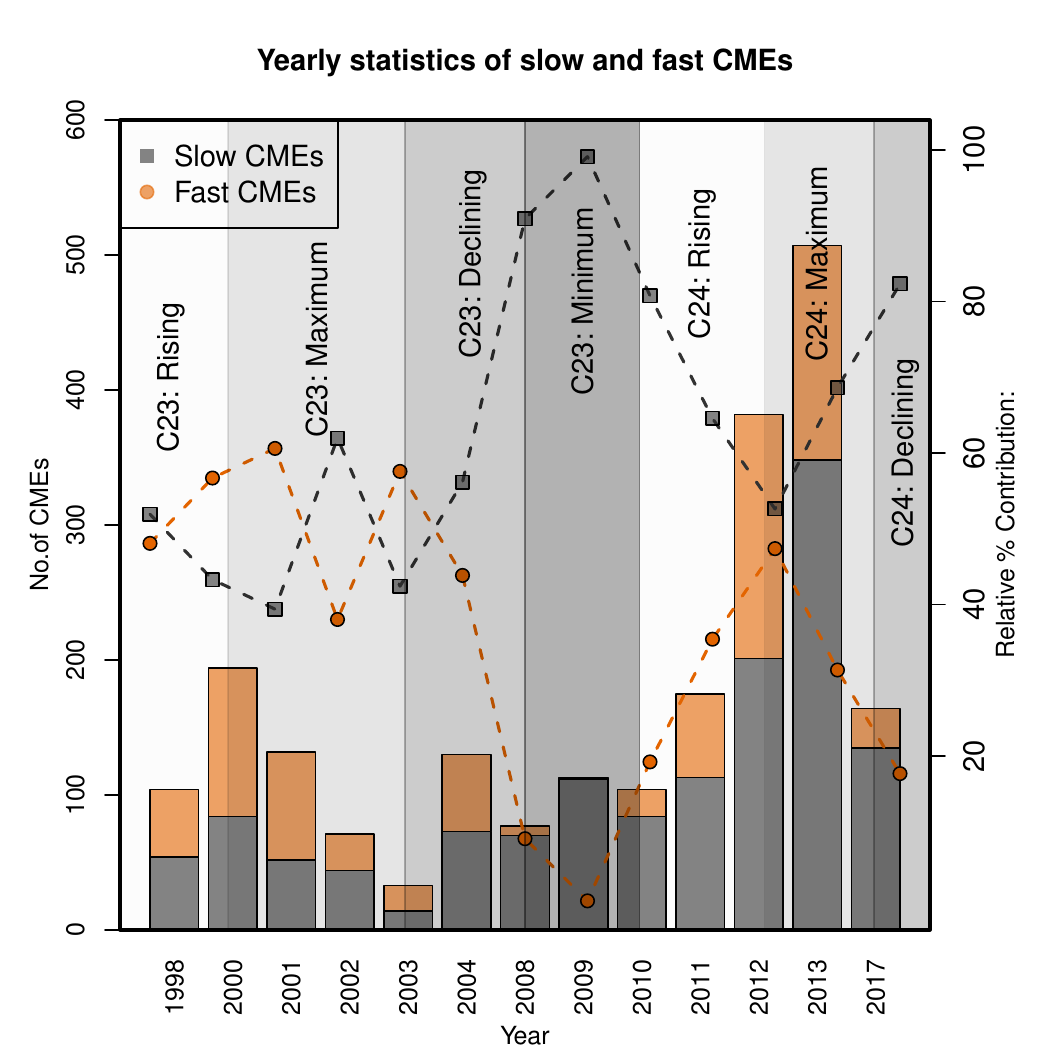}{0.5\textwidth}{(a)}
          \fig{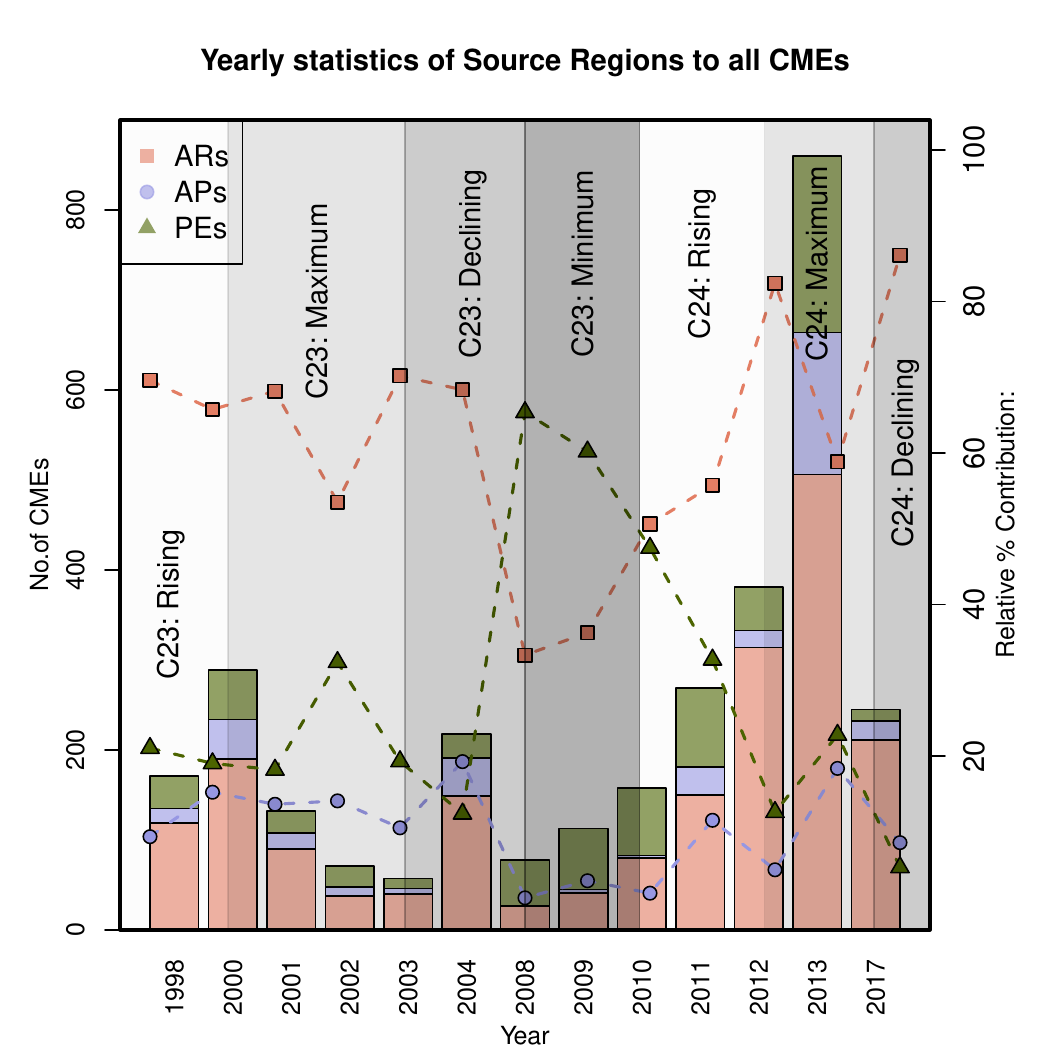}{0.5\textwidth}{(b)}
          }
\caption{This Figure shows the yearly occurrences of (a) Slow and fast CMEs, and (b) the contribution of different source regions to the CMEs. The histograms are color coded according to the category they are representing. The Y-axis on the left-hand side denote the number of events that the histograms are representing. We also show the relative contribution of each category in each year (with respect to the total number of events in that particular year) on the Y-axis plotted at the right-hand side, corresponding to the data points connected by dotted lines. The different phases of the solar cycle are highlighted in the background in different grey-scale shades.
\label{bar_speeds_sr}}
\end{figure*}

\subsection{Yearly Occurrence}

In Figure~\ref{bar_speeds_sr}(a) we plot the yearly occurrences of slow and fast CMEs during our period of study. The color-coded histograms denote the occurrences of slow and fast CMEs relative to each other for each year. On the right hand side, we have plotted another Y-axis which displays the relative percentage of occurrences of the slow and fast CMEs with respect to the total CMEs, for the ease of appreciating the differences. In (a) the squares represent the relative occurrence for slow CMEs and the circles represent the relative occurrences of fast CMEs, while in (b), the squares represents relative contribution of ARs, the circles for APs and the triangles for PEs. Please note that since the intermediate CMEs, cannot be strictly considered as either fast or slow and in this work, since we are more interested in the slow and fast CMEs, we will not be including the intermediate CMEs in the analysis. In Table~\ref{table1}, we have grouped the yearly studied CMEs into different sub-classes, according to the phase of the solar cycle the considered years correspond to \citep[see; ][]{bibhuti_2022,hathaway_2015}. It should be noted here, that we tried to be un-biased with the selection and hence contribution of any particular category of CMEs under consideration, by studying CMEs that occurred during different phases of cycle 23 and 24. Moreover, since this work relies on statistical analysis, the conclusions arrived at from the overall behaviour of a sample will not change. Based on that, we find that during the rising phase, and as the cycle progresses towards the maximum, for both cycle 23 and 24, the fractional occurrences of slow CMEs starts decreasing, and on the other hand, the fractional occurrences of the fast CMEs starts increasing. As a result, during the solar maximum, we find almost similar occurrences of slow and fast CMEs. It is also important to note here that, despite considering similar number of months in the maxima phase of cycle 23 and 24, the number of CMEs (slow and fast) as shown in the histograms, is higher in cycle 24 maxima, as compared to cycle 23 maxima, thus confirming the occurrence of more CMEs in similar period of study in cycle 23 and 24. This is supported by the report on relatively higher CME rate in cycle 24 compared to cycle 23 by \cite{lamy_2019}, where \cite{gopalswamy_2014} suggested a weakening of the overall poloidal field that led the weaker CMEs to escape out into the heliosphere. This is further reflected into the fact that the number of slow CMEs in cycle 24 always exceeds the number of fast CMEs (Figure~\ref{bar_speeds_sr}(a)). The declining phase on the other hand, shows a swinging/alternating behaviour, indicating repeating increase and/or decrease in the fractional occurrences of the slow and fast CMEs. However, as the cycle further progresses towards the minimum, we see that the fractional occurrences of the slow CMEs largely exceeds the occurrences of fast CMEs. We also see that even for solar minimum, we do not get to see many fast CMEs, but we do get to observe a reasonable number of slow CMEs. Thus, during the solar minimum, while the Sun is expected to remain quiet, in terms of CMEs, the Sun is not really quiet.

Since it is evident, that the occurrences of slow and fast CMEs vary during the different phases of a solar cycle, it also becomes imperative to look into the origin of such variations. In order to do that, we first plot in Figure~\ref{bar_speeds_sr}(b), the yearly contribution of the different source regions types (which are in our work, ARs, APs and PEs) to the CMEs (both slow and fast). We find that the contribution of the ARs dominate over the contributions from APs and PEs for most phases of the solar cycle. However, during the solar minimum of cycle 23, we find that the contribution of the ARs drop down, which is expected, but at the same time, the contribution of the PEs rise up. Thus we see that during the minimum phase, most of the CMEs are coming from the PEs, which are rooted into the quiet Sun region. The contribution of APs, is found to be comparatively lesser than those from ARs and PEs. As the cycle progresses towards the solar maximum, we can see that the number of contributions from PEs increases, but the contribution from ARs during the same increases even more, and hence that leads to a declining trend of the relative contribution from PEs.

\begin{figure*}[h]
\gridline{\fig{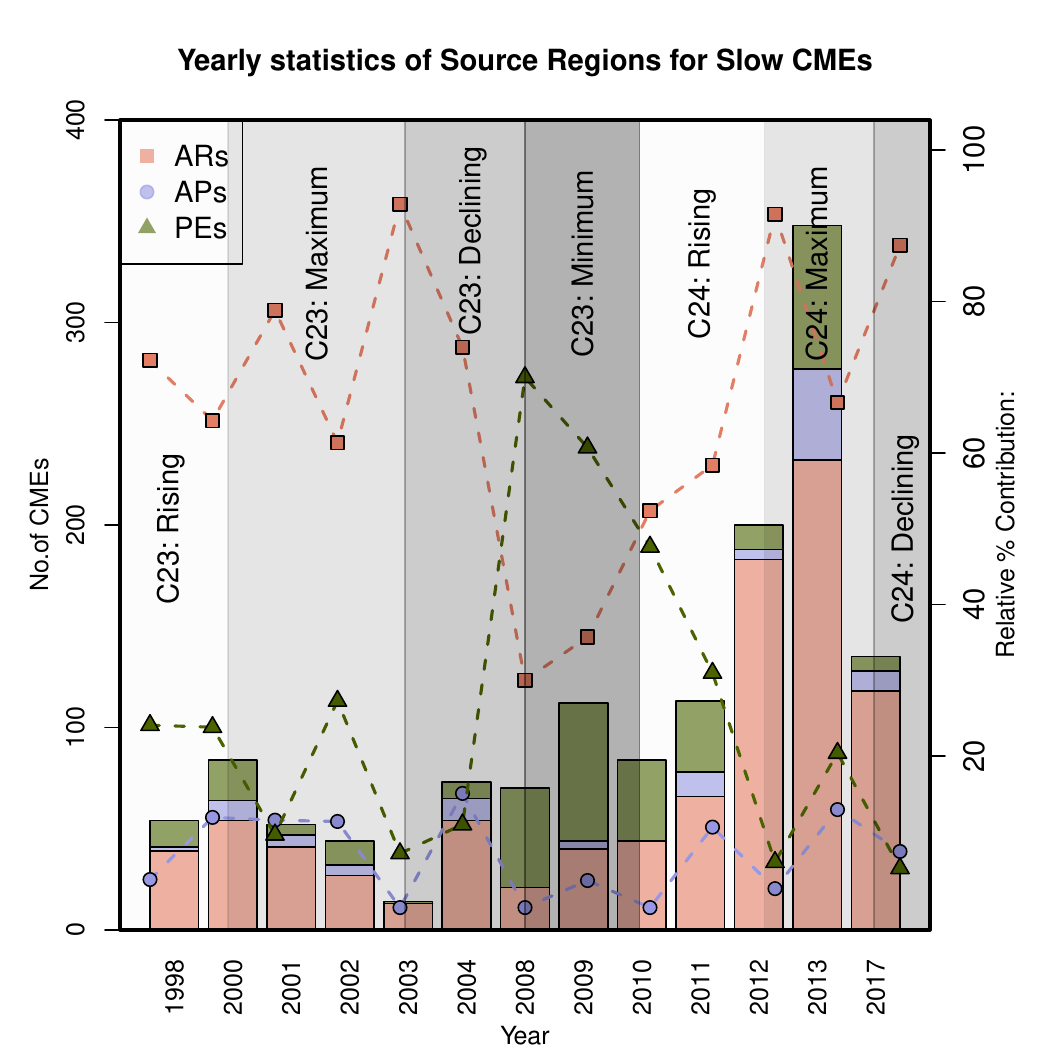}{0.45\textwidth}{(a)}
          \fig{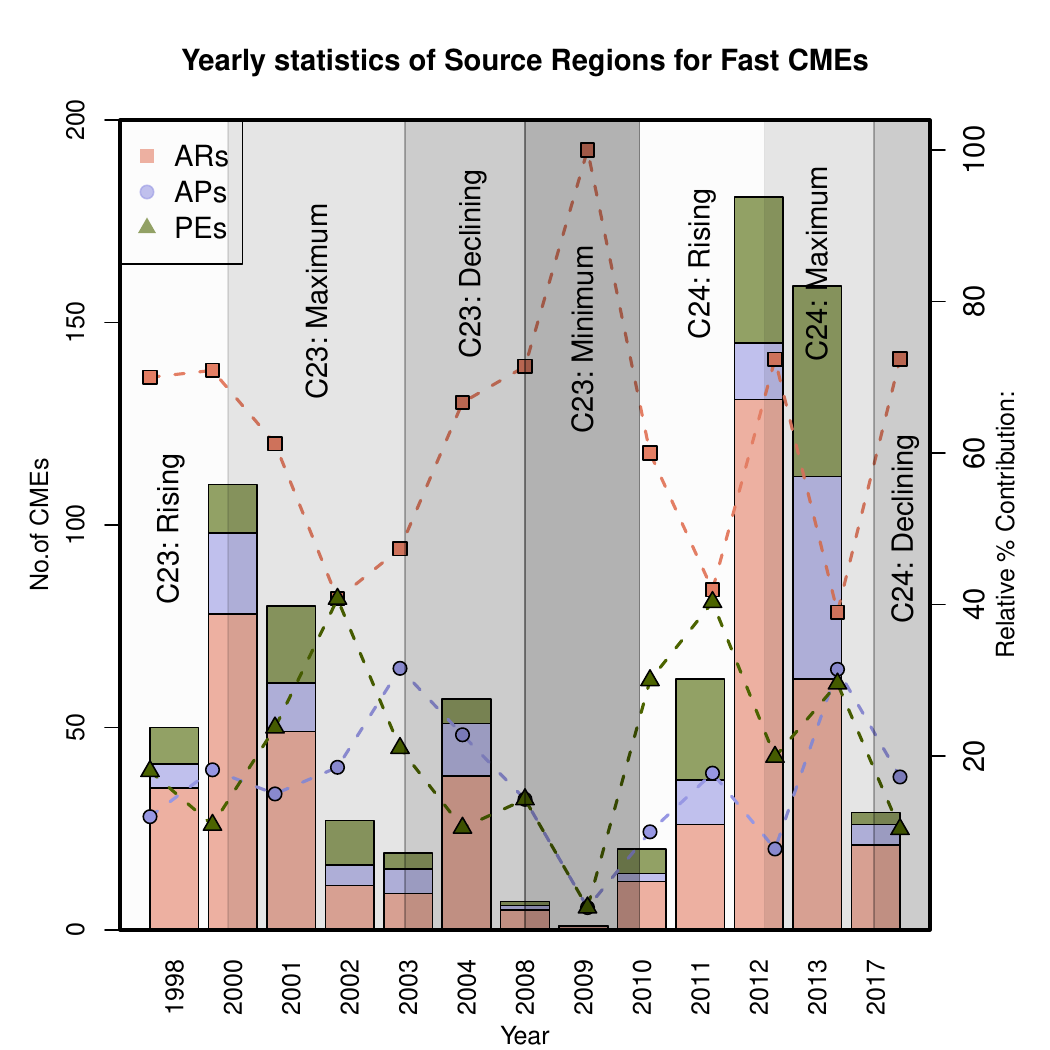}{0.45\textwidth}{(b)}
          }
\caption{Yearly contributions of source regions to (a) slow CMEs and (b) fast CMEs. The histograms are color coded according to the category they are representing. The Y-axis on the left-hand side denote the number of events that the histograms are representing. We also show the relative contribution of each category in each year (with respect to the total number of events in that particular year) on the Y-axis plotted at the right-hand side, corresponding to the data points connected by dotted lines The different phases of the solar cycle are highlighted in the background in different grey-scale shades. 
\label{bar_sr_sf}}
\end{figure*}

Having found that apart from the variation in the slow and fast CMEs across the different phases of the solar cycle, the contribution of the different source regions, also varies, it becomes crucial to combine these two results and thereby find the essence of the variations in slow and fast CMEs coming from these different source regions. To do that, in Figure~\ref{bar_sr_sf}, we look into the contribution of these source regions to slow CMEs (in (a)) and fast CMEs (in (b)) separately. In the case of slow CMEs (Figure~\ref{bar_sr_sf}(a)), we see that the contribution of ARs increases as the cycle progresses from the rising phase to the cycle maximum in both cycle 23 and 24. During the minimum in cycle 23, we see that the contribution of ARs drops down, while the contribution from PEs increases. Thus we note that most of the slow CMEs, occurring during the solar minimum in cycle 23, arise from PEs. In this context, \cite{lugaz_2017} have reported that these slow CMEs are often capable of driving shocks and hence they are of importance from the space weather perspective as well. Thus, it is important to capture and understand the variations of these slow CMEs, whose occurrences exceed the occurrences of fast CMEs during the solar minimum. As the cycle progresses further from the minimum of cycle 23 to the rising phase and then the maximum of cycle 24, we again see the contribution of ARs increase much more than the contributions from PEs, and once again, the contribution from APs is found to be much lesser than the contribution from the other two classes.  However, it is also worth noting that, a separation of APs from the other two classes have facilitated in the demarcating contributions of ARs and PEs during the different phases of the solar cycle.

In the case of the fast CMEs (Figure~\ref{bar_sr_sf}(b)), the contribution of ARs is higher than the ones from APs and PEs. Nevertheless, we also find that the contribution of APs to fast CMEs is much higher than its contribution to slow CMEs. The contribution of PEs also increases as the cycle progresses from the rising phase to the maximum phase, while during the minimum, in cycle 23, we find that the handful of fast CMEs that occur, originates mostly from ARs. Interestingly, we also note that the contribution of APs exceeds the contribution from PEs during the declining phase of cycle 23 and 24.

\begin{figure*}[h]
\gridline{\fig{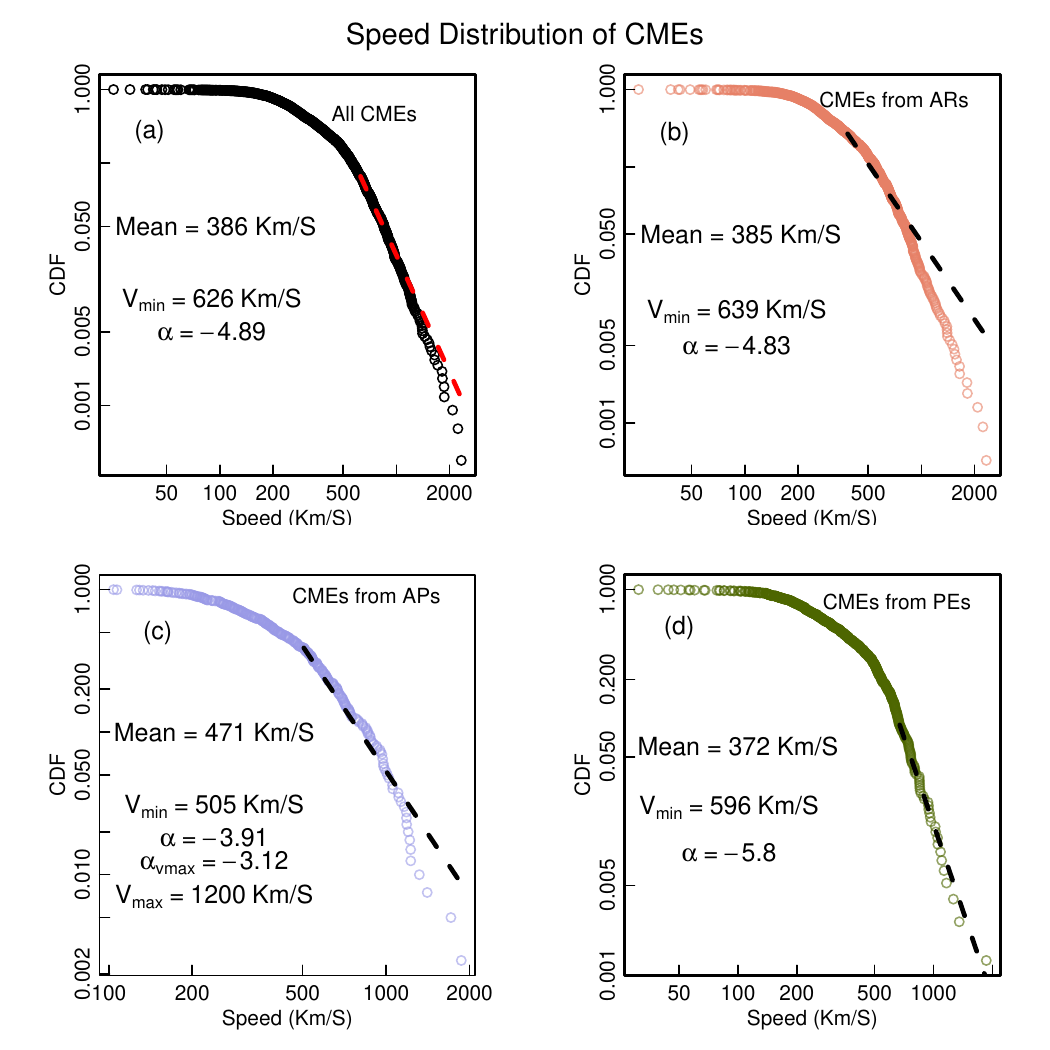}{0.9\textwidth}{}
          }
\caption{In this Figure, we plot the (d) Distribution of average speeds of all CMEs and the distribution of average speeds of CMEs coming from ARs, APs and PEs in (b), (c) and (d) respectively. A power law is fitted to the data based on MLE (in dashed line). The power law index ($\alpha$) and the mean speed is also provided in each panel.
\label{sp_pow}}
\end{figure*}

\subsection{Distribution of average CME speeds}

The average speeds have been extracted from the CDAW catalogue, which is the projected speed. In Figure~\ref{sp_pow}, we plot the distribution of the average CME speeds. We plot the cumulative distribution function here, as it is straightforward to calculate, and no binning is required, which further removes the bias that is often introduced by the choice of bin widths \citep[][]{white_2008}. In panel (a), we plot the distribution of the average speeds of all the CMEs, where the mean speed of the distribution is 386 km/s, and to that we fit a power-law. In order to get a better estimate on the fitted power-law to a discrete data-set \citep[see; ][]{d'huys_2016}, we use the Maximum Likelihood Estimate (MLE) technique to fit the power law. From Figure~\ref{sp_pow}, a minimum value of the speed (say v$_{min}$) is seen (for which the power-law behaviour is observed), and the choice of this $v_{min}$ parameter is crucial as that in turn dictates the power-law index. Thus, to find $v_{min}$, first an initial guess of $v_{min}$ is set as the minimum value of the data set. Now for a given $v_{min}$, MLE technique is used to fit a power-law to the data. Then the entire data set is scanned with different values of $v_{min}$, and finally $v_{min}$ is decided for which the Kolmogorov-Smirnov (KS) distance \citep{clauset_2009} for the fitted power-law is minimized. Recently, \cite{majumdar_2020} reported that the catalogued speeds, especially for fast CMEs can be highly misleading, as they are often observed as halo or partial-halo CMEs, and for such CMEs, there can be a reasonable overestimation of their propagation speeds, and can also be mistaken for their expansion speeds. Further, these CMEs are less frequent, and tend to occur in clusters \citep{ruzmaikin_2011}, seldom exhibiting extreme kinematic properties \citep{gopalswamy_2018}. As a result, the tail of the distribution often shows a sudden break, where the power-law behaviour is no longer followed. Thus to remove the influence of the data points occurring at this broken tail, we put an upper cut-off ($v_{max}$) to the fitted power-law, indicating that any data above $v_{max}$ is ignored for fitting. In Table~\ref{table2}, we have listed the power-law indices for all the categories discussed above,  the lower cut-off $v_{min}$, the upper cut-off $v_{max}$, their corresponding power-law indices, and the associated p-values for the statistical significance of the fitted power-laws. It can be seen that the p-values are higher than the threshold value (0.05), thus indicating their statistical significance. Please note that the upper cut-off $v_{max}$ and the corresponding $\alpha$ is provided for only those cases for which a break in the tail of the distribution is noted. We find a power law index of -$4.89$. Since we have the information of the source regions the CMEs are coming from, in panels (b), (c) and (d), we again fit a power-law separately for the CMEs originating from ARs, APs and PEs, using the same technique. In this case, we find an average speed of 385 km/s, 471 km/s and 372 km/s for the distribution of speeds of CMEs from ARs, APs and PEs respectively. We find that the average speed of CMEs from APs is higher than the average speed for CMEs from ARs. This is due to the fact that the number of slow CMEs largely exceeds the number of fast CMEs in cycle 24 (see Figure~\ref{bar_speeds_sr}), and in Figure~\ref{bar_sr_sf}, we see that most of these slow CMEs are getting contributed from the ARs. Further, \cite{gopalswamy_2020} reported on an average lower potential energy of the ARs which have resulted into a larger number of weak and slow CMEs in cycle 24. We find that CMEs originating from different source regions follow different power-law profiles. The CMEs originating from PEs follow a steeper power-law with a power-law index of -$5.8$ than the ones originating from ARs and APs, which have a power-law index of -$4.83$ and -$3.91$ respectively. Here, in the case of CMEs coming from APs, we note a break in the tail of the distribution, and thus by implementing the $v_{max}$ parameter, we find a slight change in the power law index, with the new $\alpha$ being -3.12. \cite{pant_2021} have recently shown that the power-laws associated with the width of slow and fast CMEs also show a similar trend, with the width of the CMEs originating from PEs following a steeper power-law than the ones from ARs. It seems the width and speed distribution of CMEs tend to exhibit a similar imprint of source regions on them.

\begin{figure*}[h]
\gridline{\fig{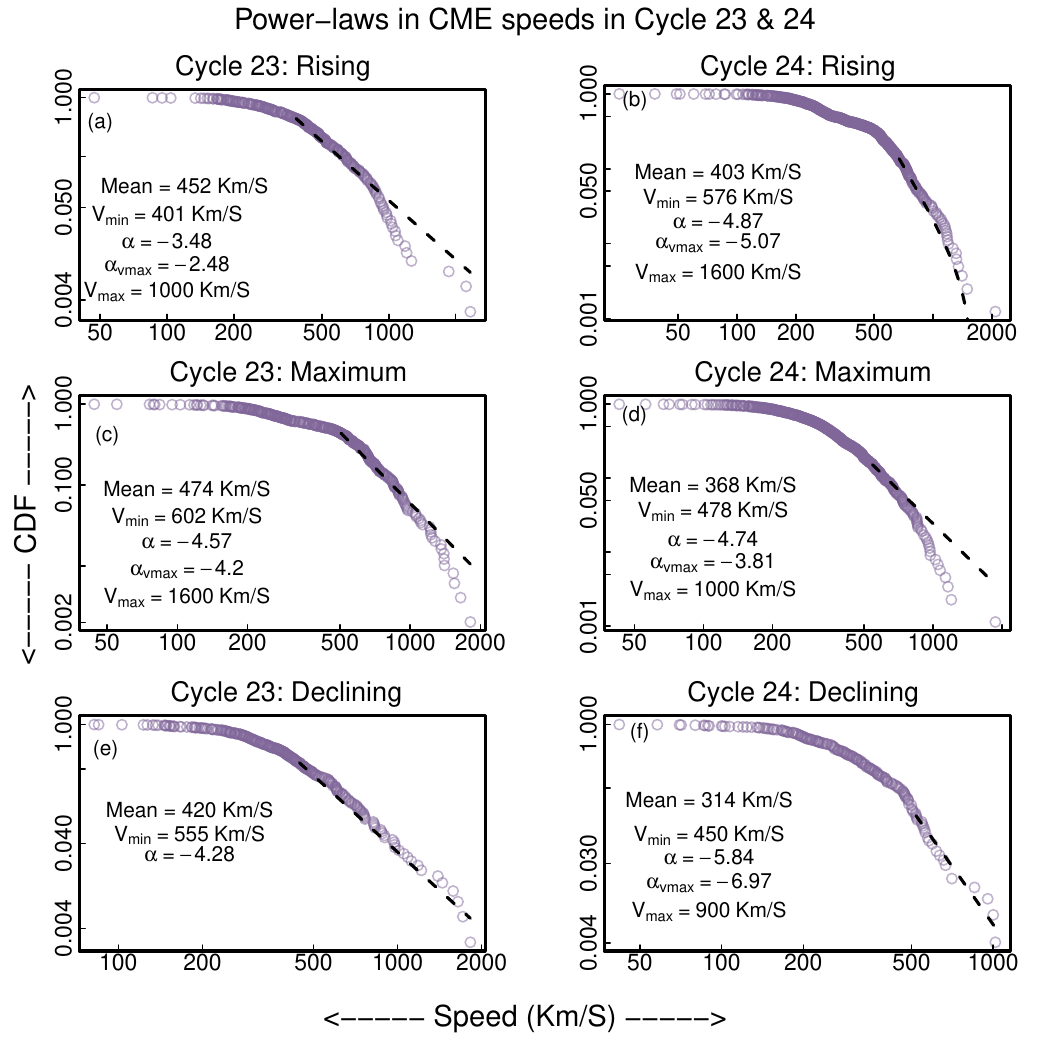}{0.85\textwidth}{}
          }
\caption{This Figure shows the speed distribution of CMEs, fitted with a power-law. (in dashed line), for different phases of the solar cycle 23 and 24. The fitted power-law index ($\alpha$) is mentioned in each case.
\label{sp_powr_cycl}}
\end{figure*}

\subsection{Variation of speed distribution with the phase of solar cycle}



In Figure~\ref{sp_powr_cycl}, we plot the distribution of the average speeds of CMEs, occurring during similar phases of cycle 23 and 24 adjacent to each other. In panel (a) and (b), we plot the distribution of the speeds during the rising phases of cycle 23 and 24. We find that the mean speed is higher in cycle 23 (467 km/s) as compared to cycle 24 (405 km/s) for the rising phase. For the maximum phase, in panel (c) and (d), we again get a similar trend. The average speed in cycle 23 maximum is 484 km/s, while that in cycle 24 maximum is much lower with a mean speed of 370 km/s.  For the decaying phase again, we find that the mean speed is higher in cycle 23 (434 km/s) as compared to the mean speed in cycle 24 (314 km/s). Thus, overall we find that the average speeds during cycle 24 are always lesser than the average speeds in cycle 23, irrespective of the phase of the solar cycle considered \citep[see; ][]{gopalswamy_2014}. Apart from that, another observable trend is that in cycle 23, the average speed is highest in the maxima, followed by the rising phase, and then the minima with the lowest average speed, while such a trend is not seen in cycle 24. In cycle 24, we find that the average speed is highest in the rising phase, followed by the rising phase and the minimum average speed in the declining phase. and we now fit a power law distribution to the data as well. For the rising phase, we find a power-law index of -$3.48$ for cycle 23 and -$4.87$ for cycle 24. For the maximum phase, we found a power-law index of -$4.37$ for cycle 23, and -$4.74$ for cycle 24, while for the declining phase, a power-law index of -$4.28$ is found for cycle 23 and a power-law index of -$5.84$ is found for cycle 24. Thus, we find that the speed distribution follow different power laws during different phases of the solar cycle. In cycle 23, we find that the maximum phase has the steepest power-law, followed by the declining phase, and with a much less steeper power-law in the rising phase. However, this is not the case for cycle 24, where we find the declining phase to have the steepest power-law, followed by the rising phase, and then the maximum phase. From Figure~\ref{sp_powr_cycl}, a break in the tail of the distribution is noted in almost all the cases, and hence a power-law is also fitted by including the upper cut-off $v_{max}$. However, we find here, that this trend remains unchanged with the inclusion of $v_{max}$ for the fitting. Although, it must also be noted that the difference in the power-law index for the rising phase and the maximum phase is much lesser compared to the difference with the other power-law indices. Apart from that, we also see that the distribution of average speeds follow a much steeper power-law for cycle 24, as compared to cycle 23, and this behaviour is independent of the phase of the solar cycle being considered.

\begin{figure*}[h]
\gridline{\fig{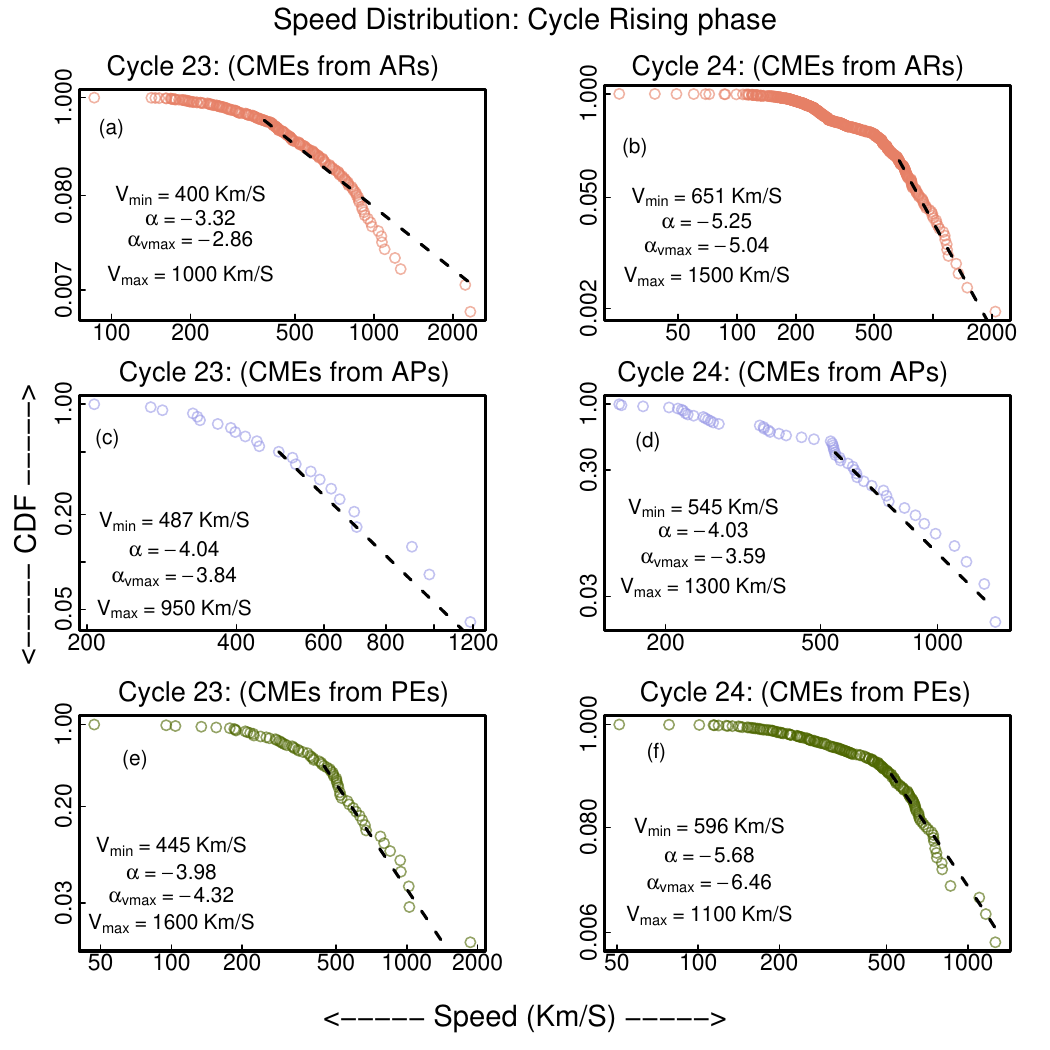}{0.85\textwidth}{}
          }
\caption{This Figure shows the variation of the speed distributions of CMEs, originating from the three different source regions (ARs, APs and PEs) and occurring during the rising phase of cycle 23 and 24. A power-law is also fitted to the data (in dashed line), and the power-law index ($\alpha$) is mentioned in each case.
\label{speed_powr_ris}}
\end{figure*}

It has now become clear that the speed distribution of CMEs follows different power laws in the different phases of cycle 23 and 24. Since we have the extra information of the source regions of these CMEs, it is also worth looking at the imprint of these source regions on this particular variation of power-laws in speed distribution across different phases of cycle 23 and 24. In Figure~\ref{speed_powr_ris}, we plot the speed distributions of the CMEs, originating from ARs, APs and PEs, and occurring during the rising phase of cycle 23 and 24. On the left panel of the plot, we include CMEs occurring in cycle 23 and on the right panel, the ones occurring in cycle 24. In cycle 23, we find a power-law index of -$3.32$, -$4.04$, and -$3.98$ for CMEs originating from ARs, APs and PEs respectively. Thus we see in cycle 23 rising phase, that the CMEs coming from APs tend follow the steepest power-law, followed by the CMEs from PEs and then from ARs. On the other hand, in cycle 24, we find the CMEs coming from PEs to have the steepest power-laws, followed by the CMEs coming from APs and then, the ones from ARs. Thus, in cycle 24, we find that the CMEs coming from PEs, follow a steeper power-law than the ones originating from ARs. This is also in agreement with the results of \cite{pant_2021}, where they had reported a steeper power-law in the width distribution of CMEs from PEs, compared to the CMEs from ARs. 

\begin{figure*}[h]
\gridline{\fig{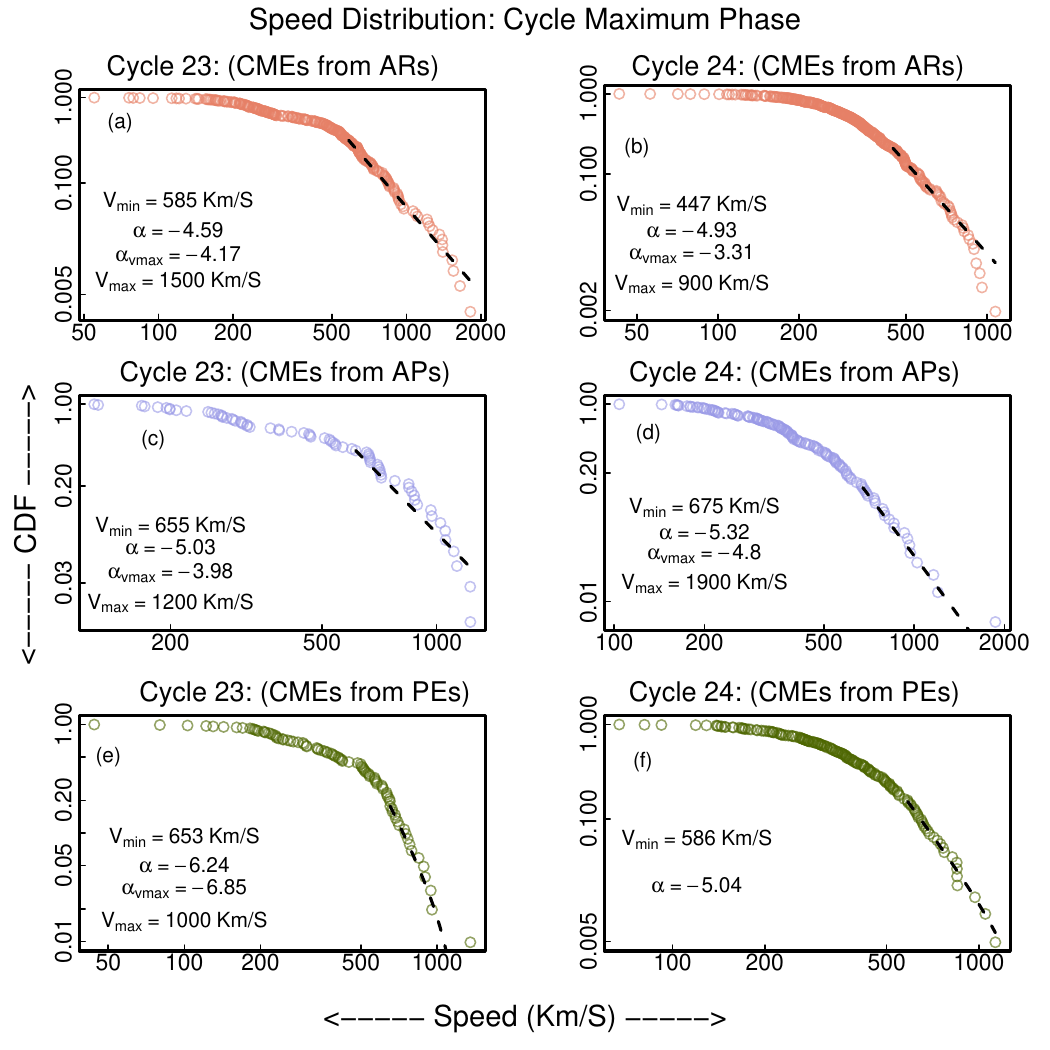}{0.85\textwidth}{}
          }
\caption{This Figure shows the variation of the speed distributions of CMEs, originating from the three different source regions (ARs, APs and PEs) and occurring during the maximum phase of cycle 23 and 24. A power-law is also fitted to the data (in dashed line), and the power-law index ($\alpha$) is mentioned in each case..
\label{speed_powr_max}}
\end{figure*}

\begin{center}
\begin{table}[]
    \centering
    \begin{tabular}{c|c|c|c|c|c}
    \hline \hline
     Average Speed for & $v_{min}$ & Power-law Index ($\alpha$) & $v_{max}$ & $\alpha$ with $v_{max}$ & P-value \\
    \hline \hline
     All CMEs & 626 & -4.89 & - & - & 0.72 \\
     CMEs from ARs & 639 & -4.83 & - & - & 0.96\\
     CMEs from APs & 505 & -3.91 & 1200 & -3.12 & 0.75\\
     CMEs from PEs & 596 & -5.80 & - & - & 0.66\\
        \hline
    All CMEs (C23 - Rising) & 401 & -3.48 & 1000 & -2.48 & 0.88\\
     All CMEs (C24 - Rising) & 576 & -4.87 & 1600 & -5.07 & 0.94\\
      All CMEs (C23 - Maximum) & 602 & -4.57 & 1600 & -4.20 & 0.87\\
     All CMEs (C24 - Maximum) & 478 & -4.74 & 1000 & -3.81 & 0.65\\
      All CMEs (C23 - Declining) & 555 & -4.28 & - & - & 0.86\\
      All CMEs (C24 - Declining) & 450 & -5.84 & 900 & -6.97 & 0.91\\
      All CMEs (C23 - Minimum) & 190 & -4.06 & 650 & -5.17 & 0.94\\
      \hline
    CMEs from ARs (C23 - Rising) & 400 & -3.32 & 1000 & -2.86 & 0.45\\
    CMEs from ARs (C24 - Rising) & 651 & -5.25 & 1500 & -5.04 & 0.76\\
    CMEs from APs (C23 - Rising) & 487 & -4.04 & 950 & -3.84 & 0.90\\
    CMEs from APs (C24 - Rising) & 545 & -4.03 & 1300 & -3.59 & 0.96\\
    CMEs from PEs (C23 - Rising) & 445 & -3.98 & 1600 & -4.32 & 0.48\\
    CMEs from PEs (C24 - Rising) & 596 & -5.68 & 1100 & -6.46 & 0.59\\
    \hline
    CMEs from ARs (C23 - Maximum) & 585 & -4.59 & 1500 & -4.17 & 0.88\\
    CMEs from ARs (C24 - Maximum) & 447 & -4.93 & 900 & -3.31 & 0.49\\
    CMEs from APs (C23 - Maximum) & 655 & -5.03 & 1200 & -3.98 & 0.87\\
    CMEs from APs (C24 - Maximum) & 675 & -5.32 & 1900 & -4.80 & 0.87\\
    CMEs from PEs (C23 - Maximum) & 653 & -6.24 & 1000 & -6.85 & 0.98\\
    CMEs from PEs (C24 - Maximum) & 586 & -5.04 & - & - & 0.64\\
    \hline
    CMEs from ARs (C23 - Declining) & 563 & -4.28 & 1500 & -4.52 & 0.41\\
    CMEs from ARs (C24 - Declining) & 444 & -5.35 & 950 & -5.55 & 0.70\\
    CMEs from APs (C23 - Declining) & 401 & -3.61 & - & - & 0.64\\
    CMEs from APs (C24 - Declining) & 254 & -3.05 & 950 & -3.19 & 0.64\\
    CMEs from PEs (C23 - Declining) & 273 & -3.03 & 700 & -2.51 & 0.31\\
    CMEs from PEs (C24 - Declining) & 137 & -2.18 & 800 & -1.61 & 0.08\\
    \hline
    CMEs from ARs (C23 - Minimum) & 157 & -3.24 & 500 & -4.39 & 0.61\\
    CMEs from APs (C23 - Minimum) & - & - & - & - & -\\
    CMEs from PEs (C23 - Minimum) & 203 & -5.99 & 800 & -6.66 & 0.92\\
           \hline \hline
    \end{tabular}
    \caption{The table shows the fitted power-law indices for the CMEs coming from different source regions during different phases of cycle 23 and 24. The minimum cut-off ($v_{min}$) for the fitted power-law is tabulated alongside. A second power-law index ($\alpha$ with $v_{max}$) is also provided in the second column for an upper cut-off ($x_{max}$, which is given in the adjacent column) at the tail of the distribution. The associated p-values for the fitted power-laws are also provided alongside.} 
    \label{table2}
\end{table}
\end{center}

Similarly, in Figure~\ref{speed_powr_max}, we plot the speed distributions of the CMEs, originating from ARs, APs and PEs, and occurring during the maximum phase of cycle 23 and 24. In this case, for cycle 23 (on the left panel), we again find that the CMEs coming from PEs to have a much steeper power-law of -$6.24$, followed by a power-law of -$5.03$ for the CMEs from APs, and then with a power-law of -$4.34$ for the ones from ARs. For cycle 24, we find a similar trend, with the steepest power-law followed by the CMEs from PEs with a power-law index of -$5.04$, followed by the CMEs from APs, with a power-law of -$5.01$, and -$4.93$ for the CMEs from ARs. Thus, the CMEs from PEs follow a much steeper power-law compared to the CMEs from ARs and APs, irrespective of the solar cycle under consideration. For the cases in which a break in the tail of the distribution is observed, we again fitted a power-law by including $v_{max}$ in the fitting process, and we find that, in some cases, there is an appreciable change in the power-law, but the trend of the power-laws remain unchanged.

\begin{figure*}[h]
\gridline{\fig{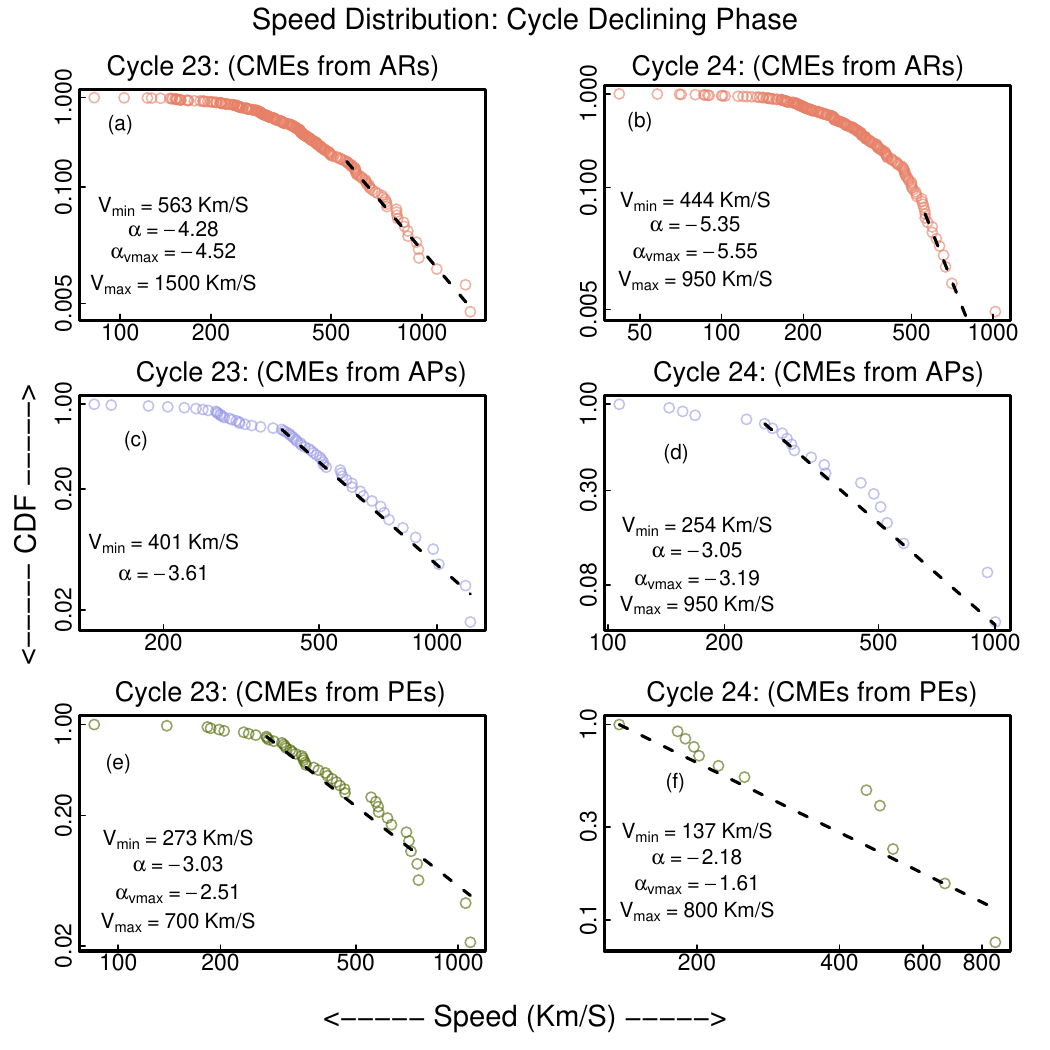}{0.85\textwidth}{}
          }
\caption{This Figure shows the variation of the speed distributions of CMEs, originating from the three different source regions (ARs, APs and PEs) and occurring during the declining phase of cycle 23 and 24. A power-law is also fitted to the data (in dashed line), and the power-law index ($\alpha$) is mentioned in each case..
\label{speed_powr_dec}}
\end{figure*}

In Figure~\ref{speed_powr_dec}, we similarly, plot the speed distributions for the declining phase of cycle 23 and 24. Contrastingly, we find that in cycle 23, the CMEs coming from ARs follow the steepest power-law, with a power-law index of -$4.28$, followed by the CMEs from APs, with a power-law of -$3.61$ and then the CMEs from PEs with a power-law of -$3.03$. A similar trend is also followed in cycle 24 as well. We find the steepest power-law of -$5.35$ to be followed by the CMEs from ARs, followed by -$3.05$ for the CMEs from APs, and then the CMEs from PEs with a power-law of -$2.18$. Although, it should be noted here, that the number of events in the class of APs and PEs in cycle 24 is lesser than the events in the class of ARs, and less number of data points may affect the derived power-laws as reported by \cite{d'huys_2016}. Having said that, it is surprising that the trend of the imprint of source regions on the distribution of speeds, reverses in the declining phase of the solar cycle, when compared to the trends set up in the rising and maximum phase. However, it must also be noted that the number of data points for the case of CMEs from PEs and APs are much lesser compared to the number of data points for the CMEs from ARs (see Figure~\ref{bar_speeds_sr}(b)), and hence in the future, by inclusion of more events for CMEs from PEs and APs would be important in understanding the trend followed in the declining phase of the solar cycle.

\begin{figure*}[h]
\gridline{\fig{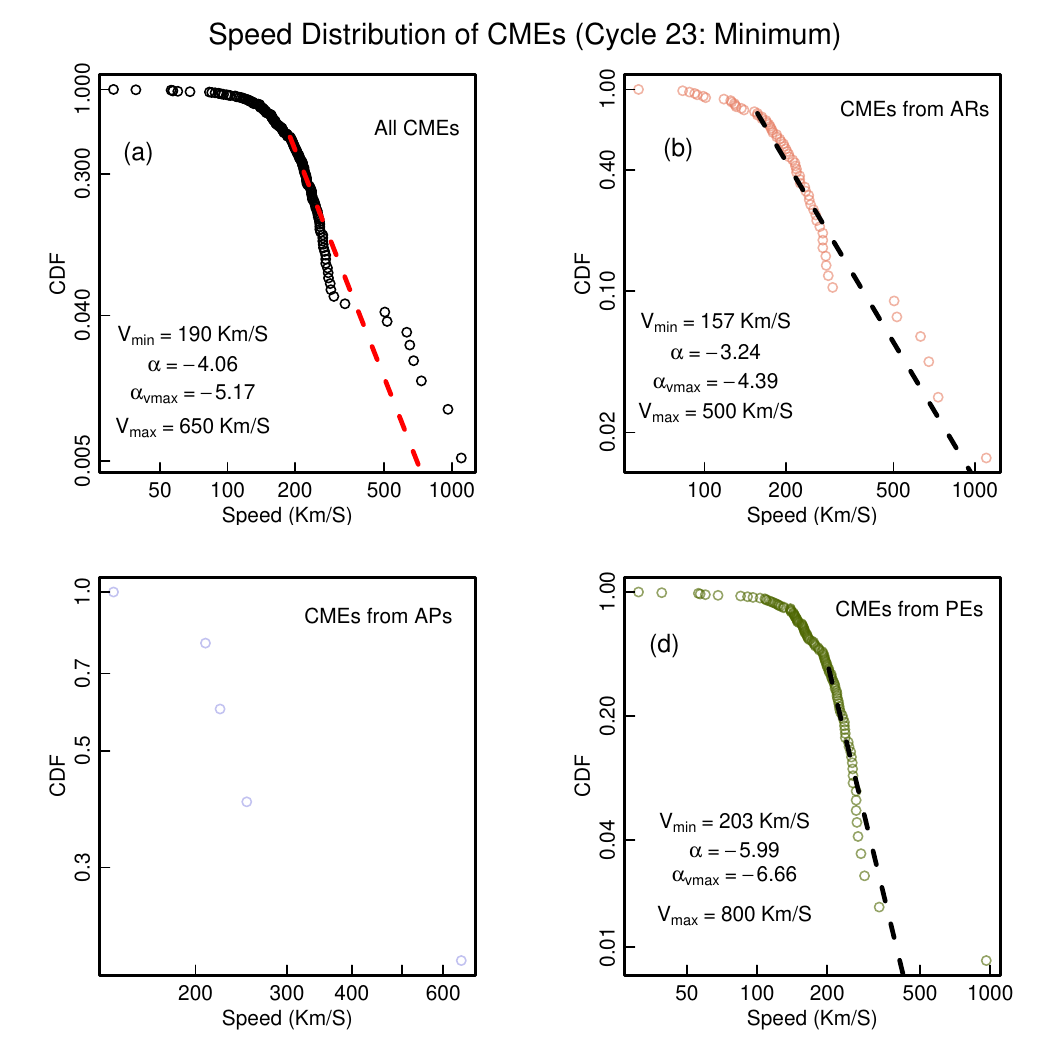}{0.85\textwidth}{}
          }
\caption{This Figure shows the variation of the speed distributions of all CMEs in (a), and those originating from the three different source regions (ARs, APs and PEs) in (b), (c) and (d) respectively, during the minimum phase of cycle 23. A power-law is also fitted to the data (in dashed line), and the power-law index ($\alpha$) is mentioned in each case..
\label{speed_powr_min}}
\end{figure*}

We study the distribution of average speeds in the minimum phase of cycle 23 in Figure~\ref{speed_powr_min}. In panel (a), we plot the distribution for all the CMEs (from different source regions), and fit a power-law to it. We find a power-law index of $-4.06$, which is lesser than what we found for the maximum phase of cycle 23 (Figure~\ref{sp_powr_cycl}). In panels (b), (c) and (d), we plot the same for the CMEs from ARs, APs and PEs respectively. Since, the number of events in the category of APs (in (c)) is much less, we were not able to fit a power-law to it. We find a power-law of $-3.24$ for CMEs originating from ARs, and a power-law of $-5.99$ for CMEs coming from PEs. Thus, in the case of the minimum phase of cycle23 too, we do see the CMEs from PEs to follow a steeper power-law in their speed distributions, with respect to the other categories. Thus, it seems the mechanisms responsible for the CMEs attaining average speeds, might be different for CMEs from PEs, compared to the ones connected to ARs, and this is independent of the phase of the solar cycle. In this case too, we notice a break in the tail of the distributions in (a) and (b), and after including $v_{max}$ for the fitting, we see that the trend of the power-laws remain unchanged.

\subsection{Latitudinal Deflections of CMEs}

A study of deflections of CMEs form an integral part of the understanding of the spatial relationship between CMEs and their source regions. In this work, we investigate the occurrences of latitudinal deflections in the CMEs considered here. In order to do that, we first convert the CPA of the CME to its corresponding equivalent latitude (with the assumption that the event is happening at the plane of the sky). This will imply that a CME with CPA of $90^{\circ}$ will have an equivalent latitude of $0^{\circ}$, and the same will also hold for a CME with CPA of $270^{\circ}$. In this context, it should be kept in mind that the CPA is measured on the plane of the sky and hence suffers from projection effects, which may lead to a conversion redundancy in the estimated latitudes, as this conversion assumes a strict radial propagation of CMEs, which is not always the case. Thus, in the future, working with the actual latitude of the CMEs found from 3D reconstruction would be crucial. A CME with a CPA of $155^{\circ}$ will correspond to a latitude of $-65^{\circ}$, which is $65^{\circ}$ South, while a CME with a CPA of $295^{\circ}$ will have an equivalent latitude of $25^{\circ}$ North. This conversion from CPA to the equivalent latitude can be generalised into the following:

\begin{equation}
    lat_{PA} = 90-CPA \quad [0 \leq CPA \leq 180] \qquad
    lat_{PA} = CPA-270 \quad [180 < CPA \leq 360].
\end{equation}

where $lat_{PA}$ is the equivalent latitude corresponding to the CPA. However, it should be noted here that the $lat_{PA}$ is a projected quantity and thus in the future, for more precise estimation of deflection, the deflection magnitude can be estimated from the actual 3D parameters \citep[see; ][]{majumdar_2020}

\begin{figure*}
\gridline{\fig{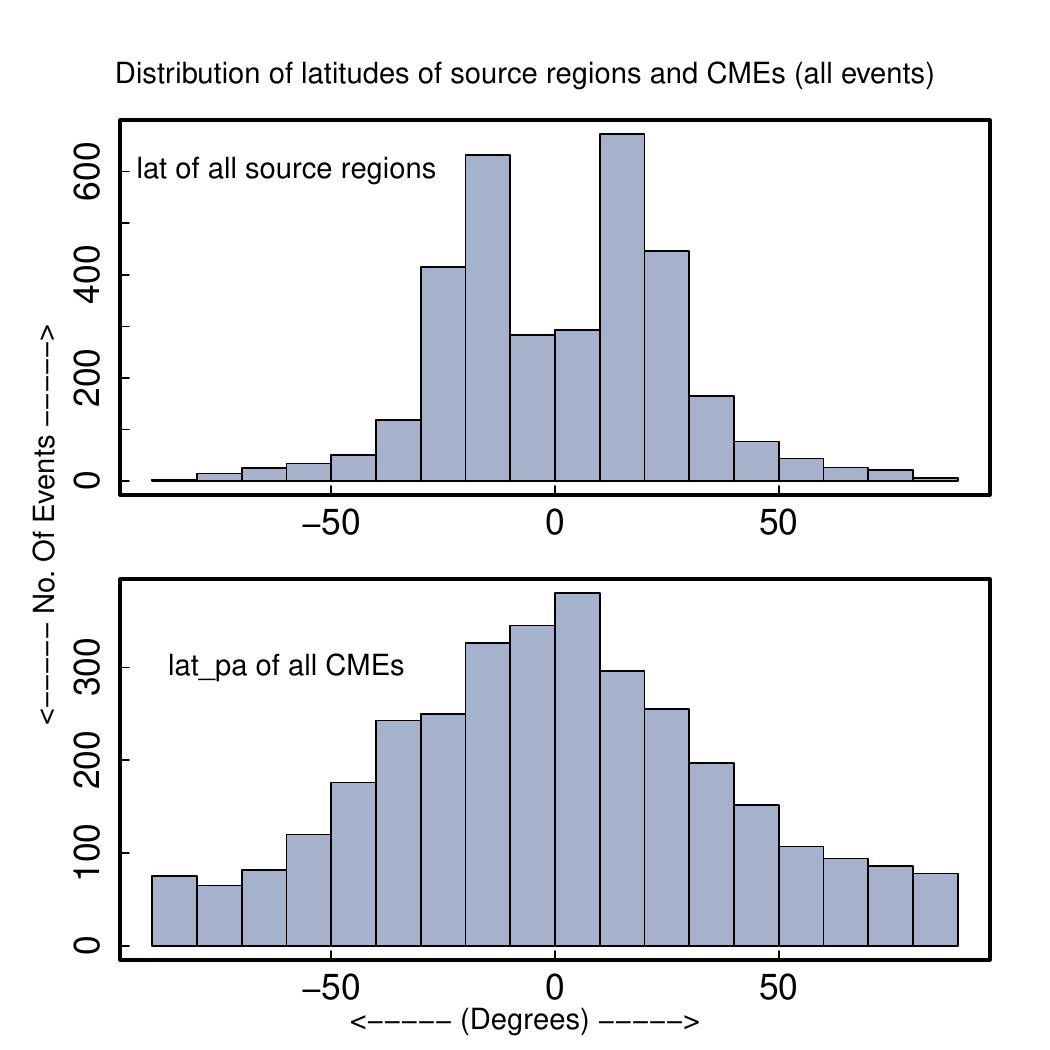}{0.39\textwidth}{(a)}
          \fig{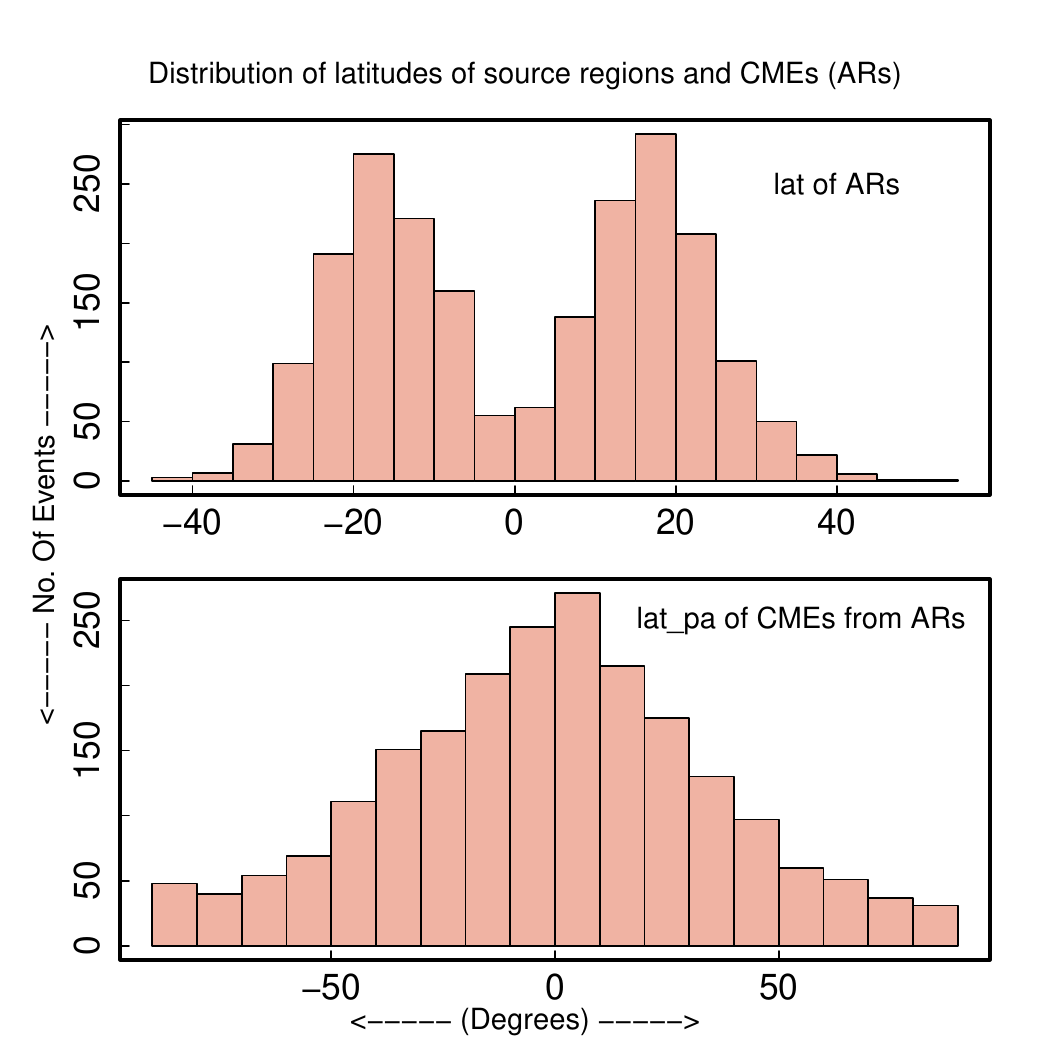}{0.39\textwidth}{(b)}
          }
\gridline{\fig{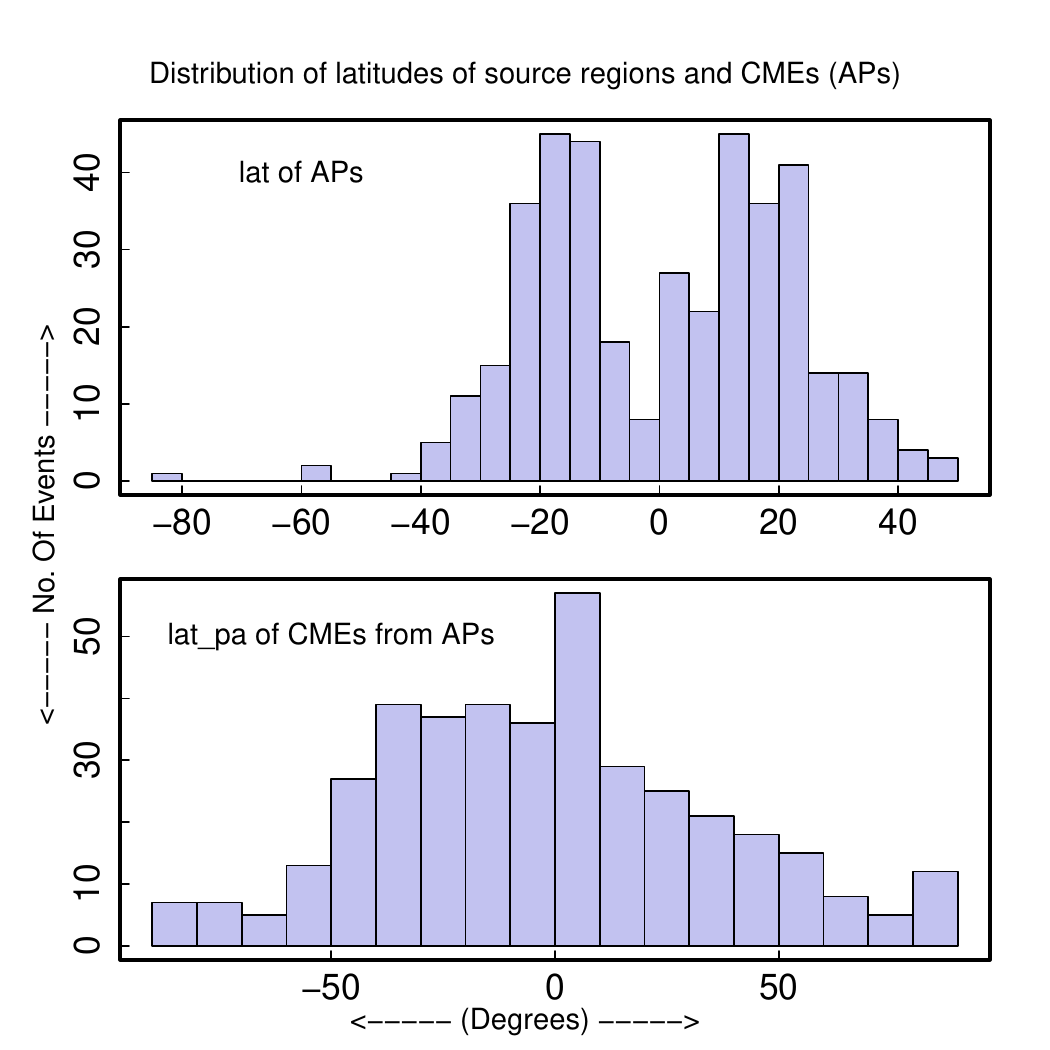}{0.39\textwidth}{(c)}
          \fig{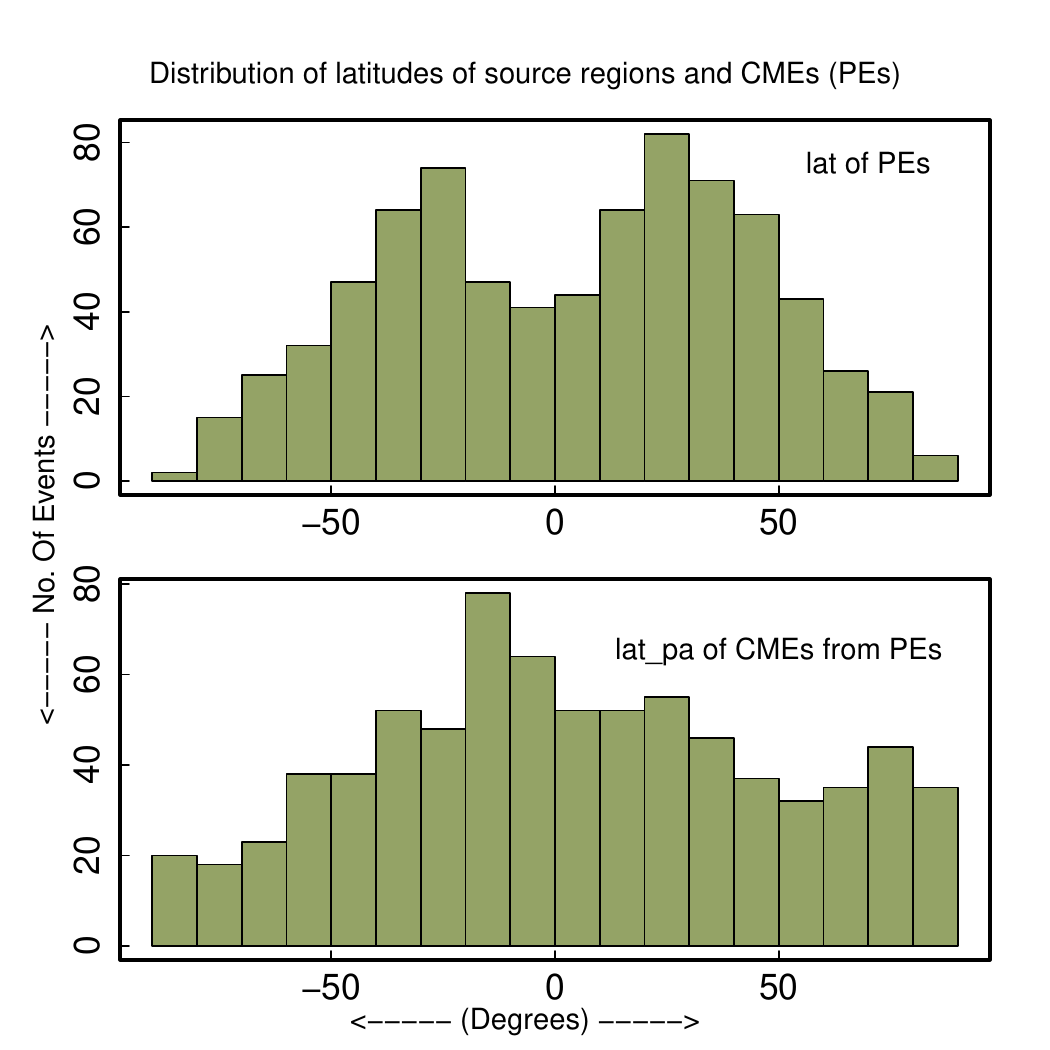}{0.39\textwidth}{(d)}
}
\caption{Latitudinal deflection of CMEs is plotted for all the CMEs in (a) and for CMEs coming from ARs, APs and PEs in (b), (c) and (d) respectively. In each figure, the top panel shows a distribution of the latitudes of the source regions, which shows a double peak signature in all the four cases. In the middle panel, a distribution of the position angle equivalent latitude is plotted in the bottom panel.\label{defl_hist}}
\end{figure*}

In Figure~\ref{defl_hist}(a), we plot the distribution of the latitudes of the source regions of all the CMEs in the top panel, followed by the distribution of the $lat_{PA}$ for all the CMEs in the bottom panel. In the top panel, we find a bi-modal distribution, with the peaks of the distribution lying at the active latitudes, around $\pm$~$10-20^{\circ}$, although, we also notice contribution from higher latitudes as well. In the distribution of $lat_{PA}$, however, we do not find a similar bi-modal distribution, we notice a single peak centered around the zero value, which is the equator, indicating that most of the CMEs, after being ejected at their respective source locations, were deflected towards the equator. A similar trend was also reported in earlier works \citep[see; ][]{gui_2011,wang_2011,cremades_2004,gopalswamy_2003} based on the association between source regions and CMEs. In our case too, since we have the source region information, we similarly plot the distributions of the latitudes of the source regions, the equivalent latitudes of the CMEs and the difference of the two latitudes, separately for CMEs coming from ARs, APs and PEs in (b), (c) and (d). We find that for all the three categories, the distribution of latitudes of the source regions are once again double peaked. While the distribution for ARs and APs peak around $10^{\circ}-20^{\circ}$, the distribution for PEs, peak around $20^{\circ}-30^{\circ}$. Further, the distribution for PEs, is more widely spread out to almost all latitudes, as was also reported in \cite{gopalswamy_2003}, while the distribution for CMEs from APs resembles that of ARs, ranging between $\pm40^{\circ}$. We also see that the distribution of $lat_{PA}$ for PEs is more spread out than the case for ARs and APs, thus indicating that the deflections, which are predominantly equatorial for the CMEs originating from ARs or APs, is not so for the CMEs coming from PEs. This behaviour might be an outcome of the fact that a reasonable fraction of PEs are located at higher latitudes.

\begin{figure*}[h]
\gridline{\fig{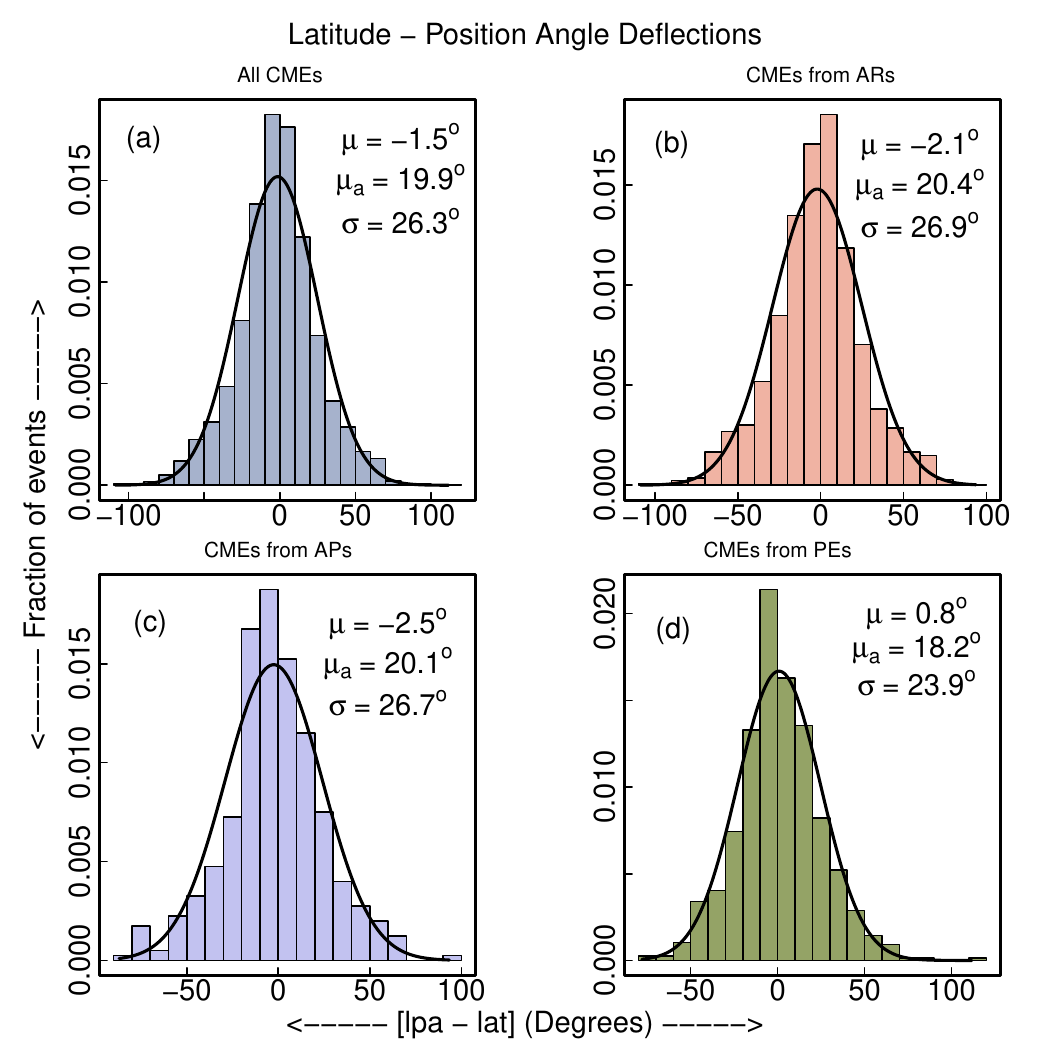}{0.8\textwidth}{}
          }
\caption{A distribution of the difference between the latitudes of the source regions of the CMEs and the position angle equivalent latitudes of the CMEs is plotted for all CMEs in (a) and for CMEs originating from ARs, Aps and PEs in (b), (c) and (d) respectively. A Gaussian is also fitted to all the four cases and the mean ($\mu$), the absolute mean ($\mu_{a}$), and standard deviation ($\sigma$) for each of the fitted Gaussian is also shown in each plot.
\label{defl_sr}}
\end{figure*}

To understand the magnitude of the deflection suffered by the CMEs, we plot in Figure~\ref{defl_sr} the distribution of the difference of the above two latitudes in each case. We find that the distribution is centered around the zero value, indicating that most of the CMEs suffered a small magnitude of deflections. This is true for the case of CMEs originating from ARs, APs and PEs as well. To have a better quantification of the same, in Figure~\ref{defl_sr}, we again plot the distribution of the difference between the source region latitudes and the CPA equivalent latitudes of the CMEs, and we fit a Gaussian to the distribution. The mean and the standard deviation of the fitted Gaussian is also shown. Further, since the deflection magnitude calculated in this way can be both positive or negative, the mean of the deflection magnitude might be misleading, and hence we also quote the mean of the absolute deflection magnitudes ($\mu_{a}$). In panels (b), (c) and (d), we separately consider CMEs coming from ARs, APs and PEs respectively. We find that the CMEs coming from PEs have the least mean (and absolute mean) deflection magnitudes of $0.8^{\circ}$ and $18.2^{\circ}$, while the CMEs from APs have a mean and absolute mean deflection magnitudes of $-2.5^{\circ}$ and $20.1^{\circ}$ respectively, and those from ARs have $-2.1^{\circ}$ and $20.4^{\circ}$. The standard deviation is similar for ARs and APs ($26.9^{\circ}$ and $26.7^{\circ}$ respectively), while it is comparatively lesser for CMEs from PEs ($23.9^{\circ}$). Thus, it seems that although CMEs from PEs exhibit signatures of deflections both towards and away from the equator (as found in Figure~\ref{defl_hist}(d)), yet the magnitude of deflections are not as spread out as the case for CMEs connected to ARs or APs. 

\begin{figure*}[h]
\gridline{\fig{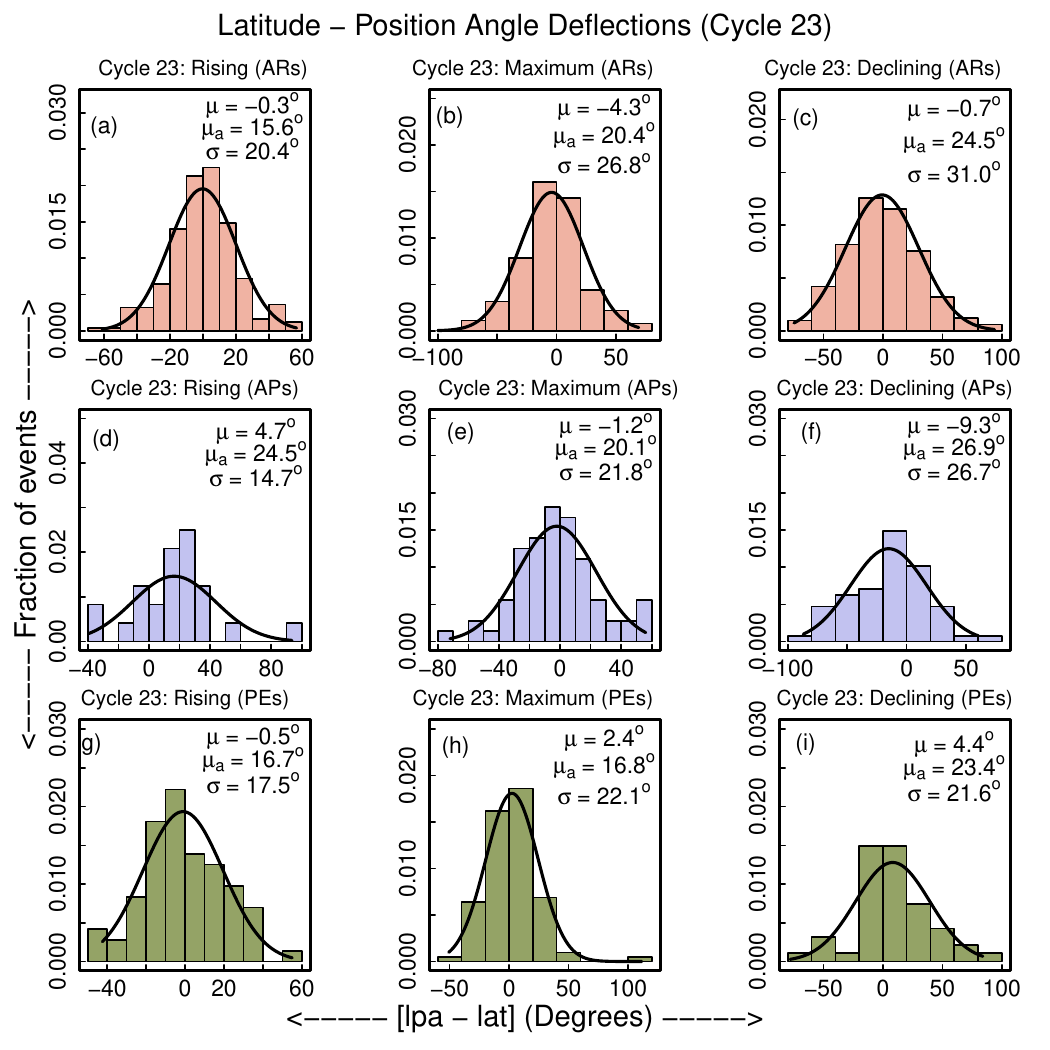}{0.99\textwidth}{}
          }
\caption{A distribution of the difference between the latitudes of the source regions of the CMEs and the position angle equivalent latitudes of the CMEs, occurring during the rising, maximum and declining phases of Cycle 23 (horizontally for each row) is plotted for CMEs originating from ARs, APs and PEs (vertically for each column, corresponding to a phase of the solar cycle). A Gaussian is also fitted to all the nine cases and the mean ($\mu$), the absolute mean ($\mu_{a}$), and standard deviation ($\sigma$) for each of the fitted Gaussian is also shown in each plot.
\label{defl_cycl23_sr}}
\end{figure*}

In Figure~\ref{defl_cycl23_sr}, we study the distribution of the magnitude of latitudinal deflection, occurring during different phases of cycle 23, and to also study if there is any imprint of source region category on the variation of this distribution. Along each horizontal row, is plotted the distribution for the rising phase, the maximum phase and the declining phase, for each of the three source region categories. While, along the each vertical column, is plotted the variation of the distribution with change in the source region category, for either of the three phases of cycle 23. We find that for CMEs connected to ARs or APs, the standard deviation is minimum during the rising phase ($20.4^{\circ}$ for ARs and $14.7^{\circ}$ for APs), it is relatively higher in the maximum phase ($26.8^{\circ}$ for ARs and $21.8^{\circ}$ for APs), while it is the highest during the declining phase, with a standard deviation of $31^{\circ}$ and $26.7^{\circ}$ for ARs and APs respectively. For PEs too, we see a similar trend, with a standard deviation of $17.5^{\circ}$ in the rising phase, to $22.1^{\circ}$ in the maximum, while a similar standard deviation in the declining phase ($21.6^{\circ}$) as that during the maximum phase. The standard deviation we find in this work is slightly higher than the overall standard deviation reported by \cite{yashiro_2009}, although they include only flaring regions under the consideration of the source regions, while not considering the phase of the solar cycle under consideration, as is done in this work. This is due to the fact that \cite{yashiro_2009} includes only those CMEs for which the longitudes of the associated flares were confined between $45^{\circ}-85^{\circ}$ degrees, while we do not put any such constraint on our events. This is reflected into a relatively lesser spread in the difference of PA distributions. 

\begin{figure*}[h]
\gridline{\fig{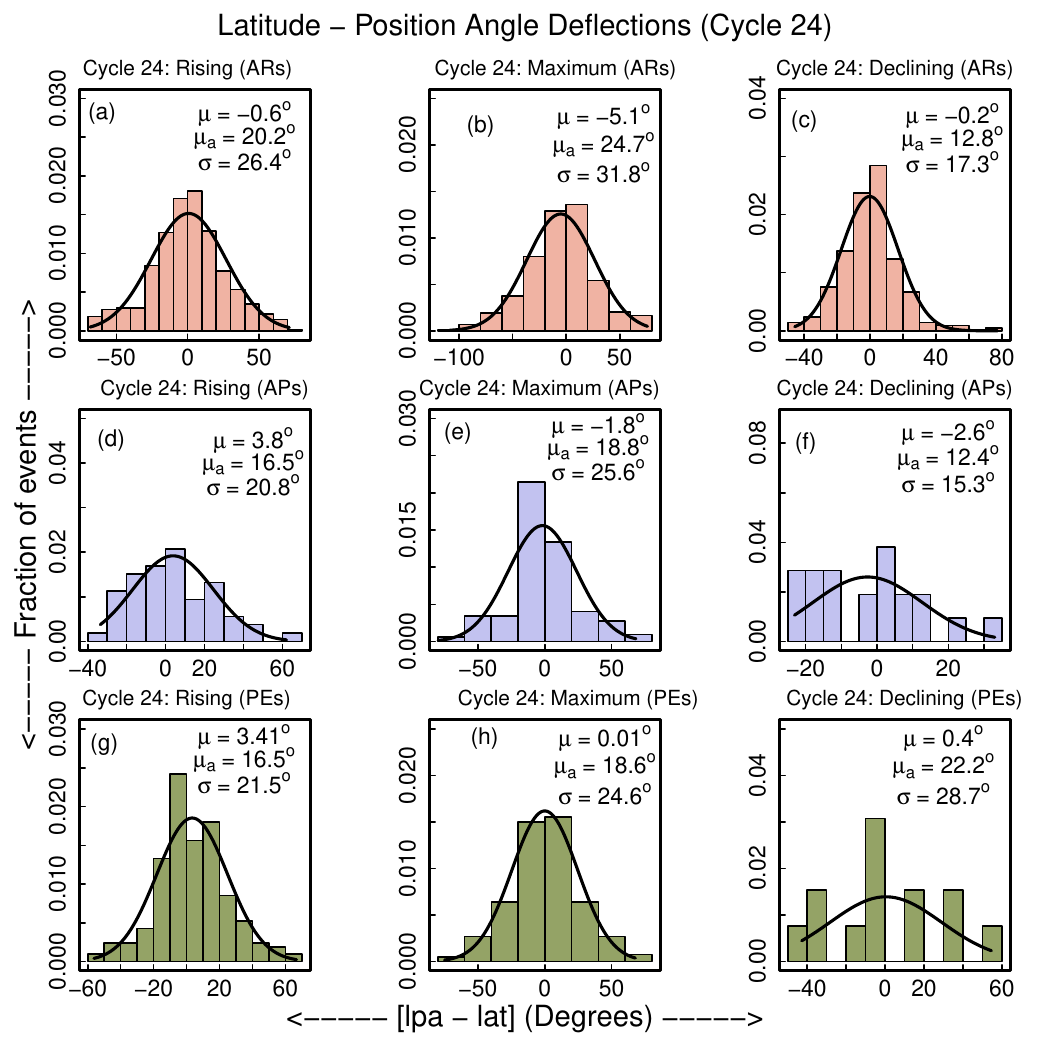}{0.99\textwidth}{}
          }
\caption{A distribution of the difference between the latitudes of the source regions of the CMEs and the position angle equivalent latitudes of the CMEs, occurring during the rising, maximum and declining phases of Cycle 24 (horizontally for each row) is plotted for CMEs originating from ARs, APs and PEs (vertically for each column, corresponding to a phase of the solar cycle). A Gaussian is also fitted to all the four cases and the mean ($\mu$), the absolute mean ($\mu_{a}$), and standard deviation ($\sigma$) for each of the fitted Gaussian is also shown in each plot.
\label{defl_cycl24_sr}}
\end{figure*}

To understand the variation of this deflection magnitude in cycle 24, we plot a similar set of figures for cycle 24 in Figure~\ref{defl_cycl24_sr}. In this case we find that for CMEs coming from ARs and APs, the standard deviation  is highest in the maximum phase ($31.8^{\circ}$ for ARs and $25.6^{\circ}$ for APs), as compared to the rising phase ($26.4^{\circ}$ for ARs and $20.8^{\circ}$ for APs)), and thereby reaching the minimum standard deviation in the declining phase ($17.3^{\circ}$ and $15.3^{\circ}$ for CMEs from ARs and APs respectively). Although we see a similar trend in case of PEs as well ($21.5^{\circ}$ in the rising phase and $24.6^{\circ}$ in the maximum phase), the standard deviation is higher in the decaying phase, but it should also be noted that the number events in the declining phase for PEs is also much lesser. Thus, it is evident that the statistical deflection trends are different during the different phases of the solar cycle, and is also dependent on the solar cycle under consideration.

\begin{center}
\begin{table}[]
    \centering
    \begin{tabular}{c|c|c|c}
    \hline \hline
    Distribution of Deflection Magnitude for & Mean deflection ($\mu$) &  Mean abs. deflection ($\mu_{a}$) & Standard Deviation ($\sigma$)\\
    \hline \hline
     All CMEs & -1.5 & 19.9 & 26.3 \\
     CMEs from ARs & -2.1 & 20.4 & 26.9\\
     CMEs from APs & -2.5 & 20.1 & 26.7\\
     CMEs from PEs & 0.8 & 18.2 & 23.9\\
        \hline
   CMEs from ARs (C23 - Rising) & -0.3 & 15.6 & 20.4\\
   CMEs from ARs (C23 - Maximum) & -4.3 & 20.4 & 26.8\\
   CMEs from ARs (C23 - Declining) & -0.7 & 24.5 & 31.0\\
   CMEs from APs (C23 - Rising) & 4.7 & 24.5 & 14.7\\
   CMEs from APs (C23 - Maximum) & -1.2 & 20.1 & 21.8\\
   CMEs from APs (C23 - Declining) & -9.3 & 26.9 & 26.7\\
   CMEs from PEs (C23 - Rising) & -0.5 & 16.7 & 17.5\\
   CMEs from PEs (C23 - Maximum) & 2.4 & 16.8 & 22.1\\
   CMEs from PEs (C23 - Declining) & 4.4 & 23.4 & 21.6\\
   \hline
   CMEs from ARs (C24 - Rising) & -0.6 & 20.2 & 26.4\\
   CMEs from ARs (C24 - Maximum) & -5.1 & 24.7 & 31.8\\
   CMEs from ARs (C24 - Declining) & -0.2 & 12.8 & 17.3\\
   CMEs from APs (C24 - Rising) & 3.8 & 16.5 & 20.8\\
   CMEs from APs (C24 - Maximum) & -1.8 & 18.8 & 25.6\\
   CMEs from APs (C24 - Declining) & -2.6 & 12.4 & 15.3\\
   CMEs from PEs (C24 - Rising) & 3.41 & 16.5 & 21.5\\
   CMEs from PEs (C24 - Maximum) & 0.01 & 18.6 & 24.6\\
   CMEs from PEs (C24 - Declining) & 0.4 & 22.2 & 28.7\\
   \hline
   All CMEs (C23 - Minimum) & -7.1 & 18.5 & 22.5\\
   CMEs from ARs (C23 - Minimum) & -8.4 & 14.9 & 17.4\\
   CMEs from APs (C23 - Minimum) & - & -\\
   CMEs from PEs (C23 - Minimum) & -5.6 & 20.4 & 25.1\\
           \hline \hline
    \end{tabular}
    \caption{The table shows the mean, the mean of the absolute deflections and the standard deviation of the Gaussian fitted to the distribution of the deflection magnitude for the CMEs coming from different source regions during different phases of cycle 23 and 24.} 
    \label{table3}
\end{table}
\end{center}

\begin{figure*}[h]
\gridline{\fig{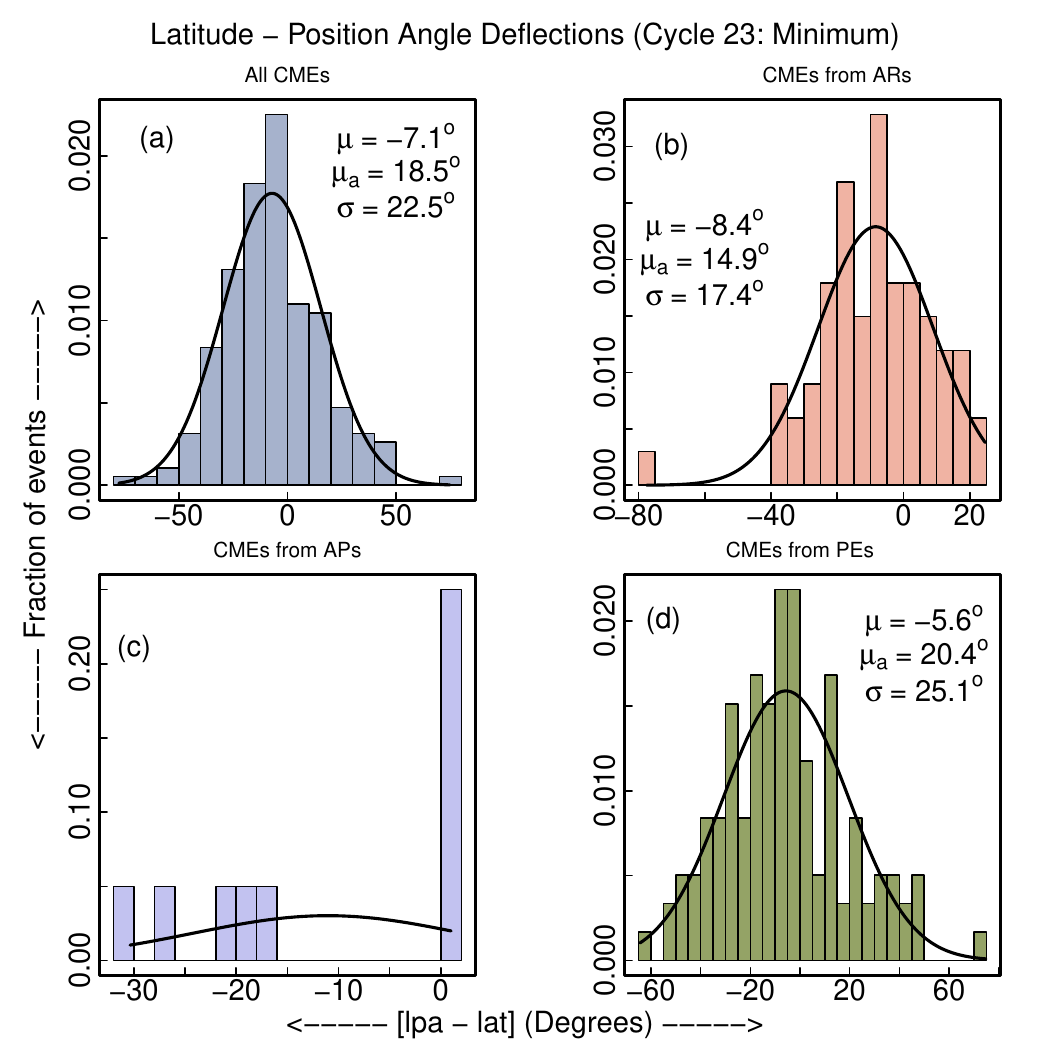}{0.85\textwidth}{}
          }
\caption{A distribution of the difference between the latitudes of the source regions of the CMEs and the position angle equivalent latitudes of the CMEs, occurring during the minimum phase of Cycle 23 is plotted for all CMEs in (a), and for the ones originating from ARs, APs and PEs in (b), (c) and (d) respectively. A Gaussian is also fitted to all the four cases and the mean ($\mu$), the absolute mean ($\mu_{a}$), and standard deviation ($\sigma$) for each of the fitted Gaussian is also shown in each plot.
\label{defl_cycl23_min}}
\end{figure*}

To understand the nature of the deflection in the minimum phase, we plot the distribution of the deflection magnitudes for the CMEs occurring during the minimum phase of cycle 23 in Figure~\ref{defl_cycl23_min}. We get an overall standard deviation of $22.5^{\circ}$ for all the CMEs (in (a)). What is worth noting in this case is the standard deviations for CMEs coming from ARs and PEs (in (b) and (d)), since for the want of statistics, we could not fit a Gaussian for the CMEs coming from APs (in (c)). For the case of ARs, the standard deviation is $17.4^{\circ}$, while for CMEs coming from PEs, the standard deviation is $25.1^{\circ}$. Thus we see that unlike the rising phase, the maximum phase or the declining phase (as shown in Figure~\ref{defl_cycl23_sr} and \ref{defl_cycl24_sr}), in the minimum phase of cycle 23, the standard deviation of the deflection magnitude is relatively higher for CMEs coming from PEs, compared to ARs. In the future, it will be worth noting if such a trend is also observed in the minimum phase of cycle 24 as well.  In Table~\ref{table3}, we have listed out the mean deflections, the mean of the absolute deflections and the standard deviations for all the categories discussed above.

\subsection{Distribution of Longitudes of Source Regions}

\begin{figure*}[h]
\gridline{\fig{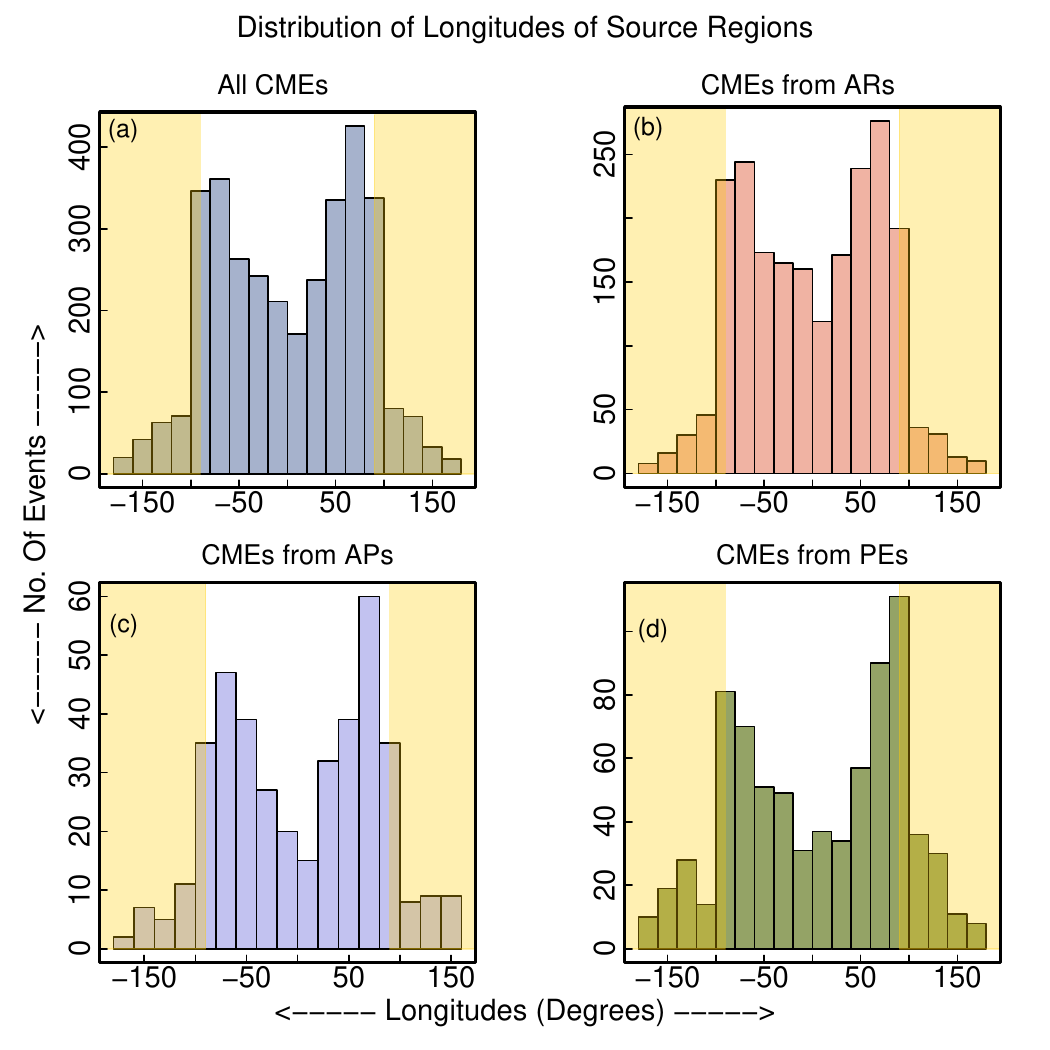}{0.85\textwidth}{}
          }
\caption{Distribution of the longitudes of the source regions is plotted for all CMEs in (a) and for the CMEs coming from ARs, APs and PEs in (b), (c) and (d) respectively. The shaded region in each plot highlights the importance of the STEREO mission in identifying and locating the source regions at the backside of the Sun (with respect to the Sun-Earth line).
\label{long_hist}}
\end{figure*}

After studying the latitudinal deflections, we study the distribution of the longitudes of the source regions in Figure~\ref{long_hist}. In panel (a), we plot the distribution of longitudes for all CMEs, and in (b), (c) and (d), we plot the same for the CMEs coming from ARs, APs and PEs respectively. In each plot, we also shade the region which highlights the significance of the contribution of STEREO mission in helping us locate and identify the source regions that were located at the back side of the Sun, with respect to the Sun-Earth line. Please note that a longitude of $0^{\circ}$ denotes the central meridian with respect to Earth. Thus longitudes lying between $90^{\circ}$ and $180^{\circ}$ signifies the back side of the Western limb, while longitudes lying between $-90^{\circ}$ and $-180^{\circ}$ signifies the back side of the Eastern limb. We find a clear double peak signature of the longitude distribution for all the four cases, (where, we find that the peaks of the distributions tend to lie in the range $\pm\,40^{\circ}-100^{\circ}$). This indicates a relatively higher contribution of CMEs originating close to the solar limb. The CMEs which originates around the central meridian are generally less prone to be detected, owing to projection effects. Further, the events studied in this catalogue is selected from the CDAW catalogue, which records CMEs detected along the Sun-Earth line. Observations from STEREO that facilitated quadrature observations, only started after 2009, so it might be possible that the relative abundance of CMEs from the limb region might be an outcome of their detection happening only along the Sun-earth line. We also find that there is a predominant East-West asymmetry in the distribution of the source regions longitudes, with a higher contribution, coming from the Western half of the Sun.

\begin{figure*}[h]
\gridline{\fig{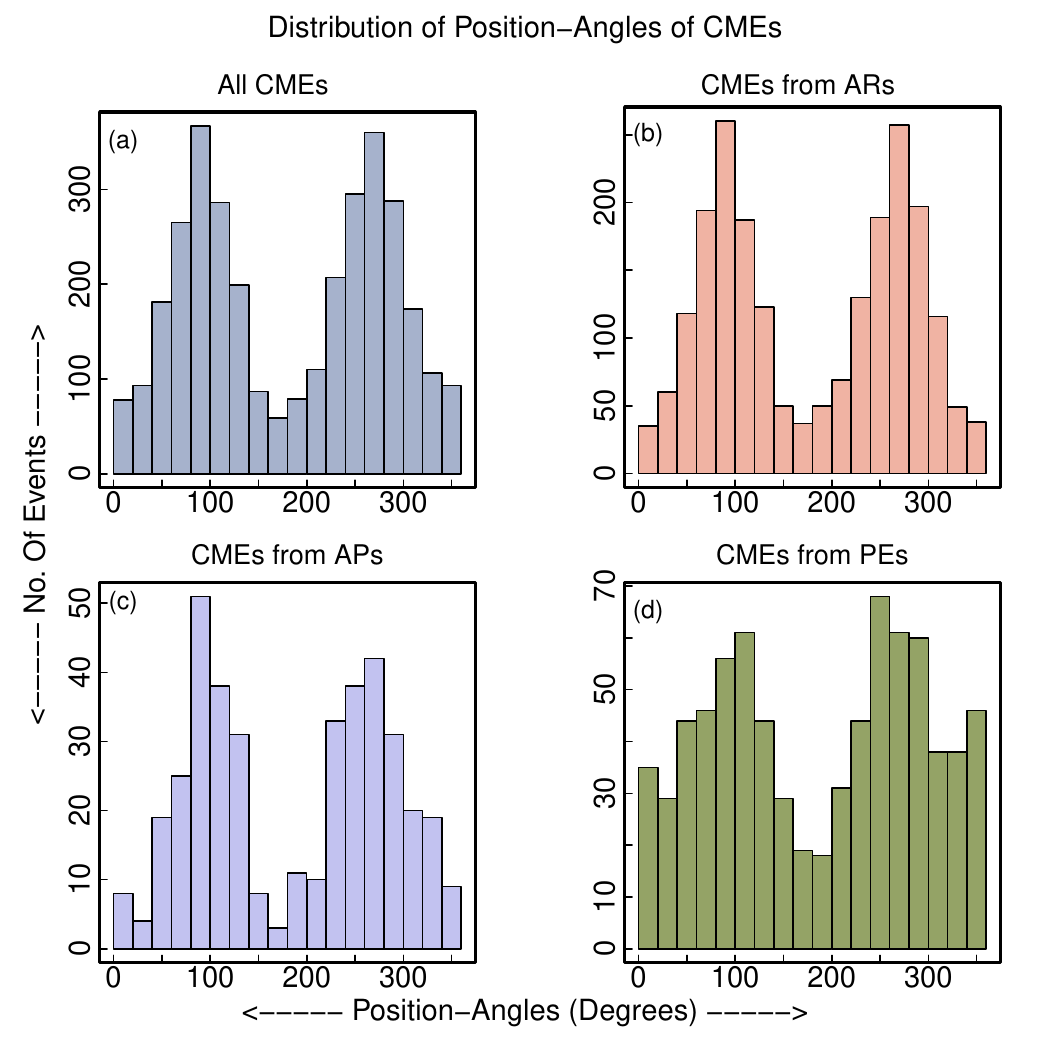}{0.85\textwidth}{}
          }
\caption{Distribution of the position angles of the CMEs, is plotted for all CMEs in (a) and for the CMEs coming from ARs, APs and PEs in (b), (c) and (d) respectively.
\label{pa_hist}}
\end{figure*}

\cite{Skirgiello_2005} reported on the existence of this East-West asymmetry in CMEs by studying the CMEs that occurred during $1996-2004$, but their conclusions were inferred from the asymmetry in the position angle of the CMEs, instead of the actual source longitudes. In order to check for the same, in Figure~\ref{pa_hist}, we plot the distribution of the position angles of the CMEs. It should be noted that in terms of the position angles, the position angles lying between $0^{\circ}$ and $180^{\circ}$ denotes CMEs along the Eastern limb, while the ones with position angles between $180^{\circ}$ and $360^{\circ}$, denote CMEs at the Western limb. We see that in this case, we do not get to see any asymmetry in the two peaks of the distribution, while it is only for the case of the CMEs coming from APs, we do see a clear East-West asymmetry, with a dominance coming from the Eastern hemisphere (which is also opposite to what we observe in Figure~\ref{long_hist}(c) suggests). Further, using the position angles to infer on the longitudinal asymmetry has its own limitations as discussed by the authors themselves in \cite{Skirgiello_2005}. Thus, studying distribution of position angles can be misleading in this context, and thus working with the actual longitudes of the source locations is encouraged.

\begin{figure*}[h]
\gridline{\fig{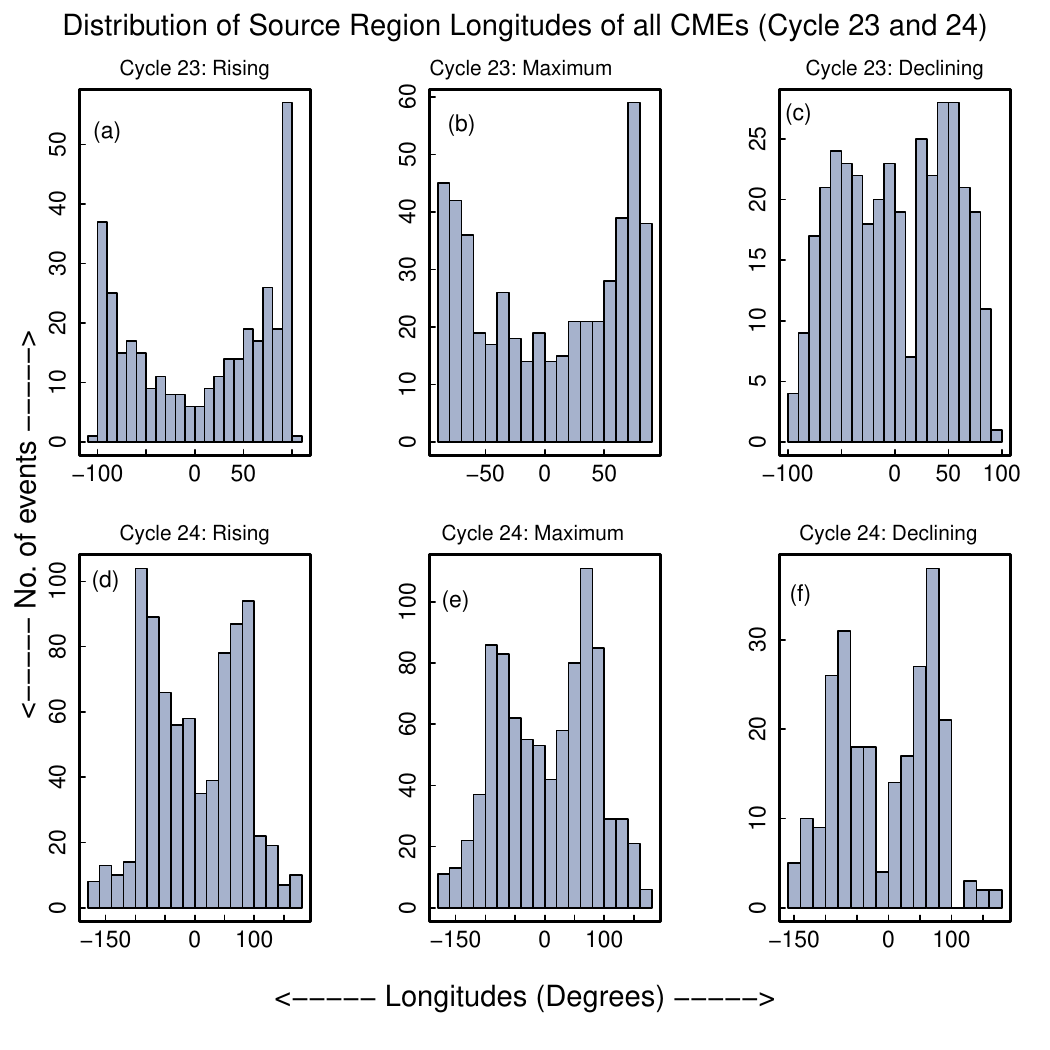}{0.95\textwidth}{}
          }
\caption{Distribution of the longitudes of all CMEs are plotted for the rising, maximum and declining phase of cycle 23 in (a), (b) and (c) respectively, and similarly for cycle 24 in (d), (e) and (f).
\label{lon_hist_cycl2324_all}}
\end{figure*}

To check whether this asymmetry undergoes through any change during the different phases of cycle 23 and 24 as considered here, we plot in Figure~\ref{lon_hist_cycl2324_all} the longitude distributions for all CMEs in cycle 23 for the rising, maximum and declining phase in (a), (b) and (c) and similarly, for cycle 24 in (d), (e) and (f). We find that only during the rising phase in cycle 23, we see a slight East-West asymmetry with a dominance from the Western hemisphere, while the distribution of the longitudes during the other phases of cycle 23 and 24 does not show signatures of any distinct East-West asymmetry. Further, this small dominance may as well be attributed to some prolific AR that spuriously appeared in the Western hemisphere and made the difference. Thus an extension of this study to a larger data set would be more insightful. However, it should be noted that \cite{buss_1918} reported that there is a preferential contribution of the prominences from the Eastern side of the Sun. Also, \cite{hey_1955} showed that there is an asymmetry in the longitudinal distribution of flares associated with radio emission, and the dominance is towards the Eastern side of the Sun. In Figure~\ref{pa_hist}(c), we find our result in support of \cite{buss_1918}, as we also observe a Eastward dominance, but again this Figure is derived from position angles, and when we look into the distribution of the actual longitudes in Figure~\ref{long_hist}(c), we no longer see the Eastward dominance. 

\section{Summary and Conclusions}\label{conclusion}

A total of 3327 CMEs are selected from the CDAW catalogue and have been studied in this work. The main aim of this work was to associate and identify each of them to the source region they are coming from, thereby cataloging the CMEs with their source regions to help understand their properties and the imprint of the source regions on the studied properties. Based on certain constraints (rejecting the "very poor" CMEs as quoted in the CDAW catalogue), after the CMEs were selected, the CMEs were segregated into slow ($<300$ km/s) and fast ($>500$ km/s) based on the average linear speed as quoted in the CDAW catalogue. Further, on the basis of certain spatial and temporal criteria, the CMEs were associated to the source regions they were coming from, thereby, classifying the identified source regions into three broad categories: (a) Active Regions (ARs), (b) Active Prominences (APs) and (c) Prominence Eruptions (PEs). Although this study does not cover all the CMEs listed in CDAW catalog (owing to constraints mentioned in Section~\ref{event_select}) for the past 25 years of LASCO observation, we have tried to create an extensive database of CMEs with their source region properties, by considering CMEs that occurred during the different phases of cycle 23 and 24. In the following, we briefly list out the main conclusions from this work:

\begin{itemize}
    \item A total of 3327 CMEs were studied out of which, 1437 CMEs were identified as slow and 786 CMEs identified as fast, while 1108 CMEs were identified as intermediate. The source regions of these CMEs were also identified, and it was found that 2037 CMEs came from ARs, 731 CMEs originated from PEs and 369 CMEs came from APs (Figure~\ref{all_cmes}).
    
    \item The CMEs selected for this study was further sub-divided into different categories (the rising phase, the maximum phase and the declining phase), based on the phase of the solar cycle, the CMEs occurred on. We found that the occurrences of slow and fast CMEs varies largely during these different phases of solar cycle (see Figure~\ref{bar_speeds_sr}(a)). We found that as the cycle progresses from the rising phase to the maximum phase, the occurrences of slow CMEs decreases, while that of fast CMEs increases, which leads to almost similar occurrences of slow and fast CMEs during the cycle maximum. On the other hand, during the minimum, we find that the number of slow CMEs largely exceeds the number of fast CMEs.
    
    \item The contribution of the different source regions also seem to be different during the different phases of the solar cycle (see Figure~\ref{bar_speeds_sr}(b)). We found that the contribution of ARs dominate over the contributions from APs and PEs for most part of the solar cycle, except during the minimum phase, during which the contribution from PEs tend to have a clear dominion over the other two.
    
    \item The contribution of the different source regions to the slow and fast CMEs also varies during our study period. We found that for slow CMEs (Figure~\ref{bar_sr_sf}(a)), as the cycle progresses from rising phase to the maximum, the contribution of CMEs from ARs increases, while during the minimum of cycle 23, we found the contribution of ARs to drop down, while the contribution from PEs increases. In the case of the fast CMEs (Figure~\ref{bar_sr_sf}(b)), we again found that the contribution of ARs dominate over the contributions from APs and PEs throughout the different phases of the solar cycle, although the relative contribution from PEs also increase from the rising phase towards the maximum phase. We also found that the contribution of APs to fast CMEs is much higher than its contribution to the slow CMEs.
    
    \item A distribution of average linear speeds of the CMEs was studied and a power-law (with a lower threshold $v_{min}$) was also fitted to the speed distribution (Figure~\ref{sp_pow}), and we found different power laws for the speed distribution of CMEs from different source regions. The CMEs originating from PEs follow a much steeper power-law, with a power-law index of $-5.8$, compared to the power-laws in the case of CMEs from ARs and APs (with power-laws of $-4.8$ and $-3.9$ respectively). It should be noted that the estimated power-law index might get influenced by the choice of $v_{min}$. Thus in this work, $v_{min}$ is calculated by KS distance minimization, and is not controlled manually, as manually constraining $v_{min}$ will introduce subjectivity bias in our analysis.
    
    \item We plot the distribution of average speeds during different phases of cycle 23 and 24 (Figure~\ref{sp_powr_cycl}), and found that the average speed in cycle 24 was lesser than the average speeds in cycle 23, irrespective of the phase of the solar cycle under consideration. We also found that the speed distributions during the different phases in cycle 23 and 24, follow different power-laws ((Figure~\ref{sp_powr_cycl}). The speed distributions in cycle 24 tend to follow a steeper power-law than the case for cycle 23, and this is independent of the phase of the cycle being compared. We also found the maximum phase to have the steepest power-law in cycle 23, and the declining phase to have the steepest power-law in cycle 24. In this context, it must be noted that since we are working with projected speeds, the estimated power-laws will be suffering from projection effects. However, from Figure~\ref{long_hist}, it is evident that most of the CMEs are originating from the limb of the Sun, and thus the projection effect corrections would be minimal for most of the CMEs considered in this work.
    
    \item We found a strong imprint of source regions on the power-laws in speed distributions in different phases of cycle 23 and 24. In the rising phases of cycle 23 and 24 (Figure~\ref{speed_powr_ris}), we found the CMEs coming from PEs to follow a much steeper power-laws than the CMEs from ARs. This trend is also followed in the maximum phases of cycle 23 and 24 (Figure~\ref{speed_powr_max}) and the minimum phase of cycle 23 as well (Figure~\ref{speed_powr_min}), where the CMEs from PEs follow the steepest power-laws. However, during the declining phases of the cycles (Figure~\ref{speed_powr_dec}), we find the CMEs originating from ARs to follow the steepest power-laws. It is crucial to note that in certain cases, a break in the tail of the distribution was noted, and to remove the influence of those data points on the fitted power-law, an upper cut-off $v_{max}$ was also included. We found that although in some cases, we do find appreciable change in the power-laws, yet the overall trends and conclusions remain unchanged. 
    
    \item A study of the latitudinal deflection of the CMEs revealed a preferential equatorial deflection of the CMEs (Figure~\ref{defl_hist}). It is important to note here, that the mean deflection might be misleading, as deflection can be both positive as well as negative. Thus quoting the mean of the absolute deflections will provide a better picture. Fitting a Gaussian to the distribution of the magnitude of deflection from different source regions, it was found that the CMEs connected to ARs and APs tend to have a wider spread in the deflection magnitudes compared to the CMEs from PEs (Figure~\ref{defl_sr}). However, during the minimum phase of cycle 23 (Figure~\ref{defl_cycl23_min}), we found that the deflection magnitude has a higher standard deviation for CMEs coming from PEs, as compared to the CMEs from ARs. We further found that in cycle 23 (Figure~\ref{defl_cycl23_sr}), the standard deviation in the distribution of the magnitude of deflection is minimum during the rising phase, and maximum during the declining phase. This trend is independent of the source region from which the CMEs are originating. On the other hand, in cycle 24 (Figure~\ref{defl_cycl24_sr}), we find the standard deviation to be lesser in the declining phase, and being highest in the maximum phase. Thus, it seems, the trend of deflections are different during the different phases of the solar cycle, and is also dependent on the solar cycle under consideration.
    
    \item A study of the distribution of the longitudes of the source regions showed a double peaked distribution, with an apparent East-West asymmetry, noted particularly in the rising phase of cycle 23, with a Westward dominance (Figure~\ref{long_hist}). A plot of the distribution of the position angle of the CMEs, although showed a similar bi-modal distribution (Figure~\ref{pa_hist}), yet concealed the asymmetry in the distribution, as was evident from the distribution in the longitudes, thus indicating that concluding based on the distribution of position angles might not give the true picture and can be misleading. This apparent asymmetry, can be a reflection of the fact that there is an abundance of limb CMEs, as CMEs originating around the central meridian are generally less prone to get detected. Besides, the data-set used in this study is taken from the CDAW catalogue, which is based on observations taken along the Sun-earth line. Also, analysing asymmetry based on the difference in heights of the tallest columns might be misleading, as the observed difference in the heights of the tallest columns may get contributed by some prolific ARs that may appear in one of the hemispheres to create the difference. Thus in the future, an extension of this study to an even larger data-set, and to CMEs detected from multiple vantage points \citep{vourlidas_2017} and based on un-projected quantities would be important.
\end{itemize}

Thus, this CSR catalogue, apart from providing a catalogue of CME containing the source region information for a sample set of 3327 CMEs, we believe our results would also implant the strong claims that the source regions have in governing several statistical properties of CMEs. However, there are certain limiting aspects of this study, as discussed above, which mainly arise from the fact that the geometrical and kinematic parameters extracted from the CDAW catalogue suffers from projection effects and from the fact that the CMEs are recorded, based on their detection only along the Sun-Earth line. Another aspect worth noting is that, not all CMEs occurring during the different phases of cycle 23 and 24 was included in this study, owing to limitations discussed in Section~\ref{event_select}. In the future, the results from this study can thus, be further extended by considering all CMEs covering the entire cycles of 23 and 24, using multiple vantage point CME catalogue. Having said that, we also believe that these results will encourage in dedicating future works on understanding the difference in the generation and propagation processes for CMEs coming from these different sources, and on the significance of the contribution of statistical works on CME behaviours, thereby, getting ourselves another step closer towards more improved space weather predictions.

\begin{acknowledgments}

We thank the anonymous reviewer for his/her careful and insightful comments that have helped in improving the manuscript. The SOHO/LASCO data used here are produced by a consortium of the Naval Research Lab- oratory (USA), Max-Planck-Institut für Aeronomie (Germany), Laboratoire d’Astronomie (France), and the University of Birmingham (UK). SOHO is a project of international cooperation between ESA and NASA. We used the CDAW catalogue data. This CME catalog is generated and maintained at the CDAW Data Center by NASA and The Catholic University of America in cooperation with the Naval Research Laboratory. SOHO is a project of international cooperation between ESA and NASA. The SECCHI data used here were produced by an international consortium of the Naval Research Laboratory (USA), Lockheed Martin Solar and Astrophysics Lab (USA), NASA Goddard Space Flight Center (USA), Rutherford Appleton Laboratory (UK), University of Birmingham (UK), Max-Planck-Institut for Solar Sys- tem Research (Germany), Centre Spatiale de Liège (Belgium), Institut d’Optique Théorique et Appliquée (France), Institut d’Astrophysique Spatiale (France). We also acknowledge the SDO team for making the AIA data available. SDO is a mission for NASA’s Living With a Star (LWS) program. The authors acknowledge Kavya Bindu M, for her contribution in preparing this catalogue. 
\end{acknowledgments}

%

\vspace{5mm}




\appendix

\section{Appendix information}



\bibliographystyle{aasjournal}



\end{document}